\documentclass[10pt,journal,compsoc]{IEEEtran}
\usepackage[noadjust]{cite}
\usepackage{authblk}
\usepackage[utf8]{inputenc}
\usepackage[T1]{fontenc}
\usepackage[
colorlinks,
    linkcolor=blue,
    citecolor=blue,
    pdftex,
    pdfauthor={assert inspired by rick},
    pdftitle={RICK with no roll},
    pdfsubject={generation of mocks with real test data and real object interactions},
    pdfkeywords={tests, execution, monitoring, rick, roll, sweat, oracles, execution},
    pdfcreator={gru}]{hyperref}
\usepackage[dvipsnames,table]{xcolor}
\usepackage{xcolor}
\usepackage{booktabs}
\usepackage{algorithm}
\usepackage{makecell}
\usepackage{multirow}
\usepackage{adjustbox}
\usepackage{balance}
\usepackage[inline]{enumitem}
\usepackage{algorithmic}
\usepackage{graphicx}
\usepackage{caption}
\usepackage{hhline}
\usepackage{tabularx}
\usepackage[scaled]{beramono}
\usepackage{xspace}
\usepackage{pgfplots}
\usepackage{subcaption}
\usepackage{cleveref}
\usepackage{soul}
\usepackage[normalem]{ulem}
\useunder{\uline}{\ul}{}
    
\pgfplotsset{compat=1.15}
\usetikzlibrary{patterns}
\usepackage{listings}
\usepackage[framemethod=tikz]{mdframed}

\mdfdefinestyle{mpdframe}{
    frametitlebackgroundcolor   =teal!15,
    frametitlerule              =true,
    roundcorner                 =1pt,
    middlelinewidth             =1pt,
    innermargin                 =0.1cm,
    outermargin                 =0.1cm,
    innerleftmargin             =0.1cm,
    innerrightmargin            =0.1cm,
    innertopmargin              =0.1cm,
    innerbottommargin           =0.1cm
}

\definecolor{light-gray}{gray}{0.95}
\definecolor{pgreen}{RGB}{5,205,107}
\definecolor{pblue}{RGB}{2,154,223}
\definecolor{ppurple}{RGB}{102,51,170}
\definecolor{ppink}{RGB}{255,20,147}
\definecolor{pteal}{RGB}{0,128,128}

\newcommand{\lstbg}[3][0pt]{{\fboxsep#1\colorbox{#2}{\strut #3}}}
\definecolor{light-gray}{gray}{0.95}
\definecolor{pgreen}{RGB}{5,205,107}
\definecolor{pblue}{RGB}{2,154,223}
\definecolor{codegreen}{rgb}{0,0.6,0}
\lstdefinelanguage{diff}{
    basicstyle=\ttfamily\footnotesize,
    language=java,
    breaklines=true,
    breakatwhitespace=true,
    keywordstyle=\bfseries\color{pblue},
    stringstyle=\bfseries\color{pteal},
    morecomment=[f][\color{red}]{---}, 
    morecomment=[f][\color{codegreen}]{+++},
    morecomment=[f][\lstbg{red!20}]{- },
    morecomment=[f][\lstbg{green!20}]{+ },
    morecomment=[f][\color{blue}]{@@},
}

\newcolumntype{R}{>{\raggedleft\arraybackslash}X}
\newcolumntype{?}{!{\vrule width 1.5pt}}

\newcommand*\circled[1]{\tikz[baseline=(char.base)]{
            \node[shape=circle,draw,inner sep=2pt] (char) {#1};}}

\newcommand{\revisedthree}{\textcolor{black}}
\newcommand{\revisedfour}{\textcolor{black}}
\newcommand{\revisedfive}{\textcolor{black}}

\newcommand{\pankti}{\textsc{pankti}\xspace}
\newcommand{\rick}{\textsc{rick}\xspace}
\newcommand{\pdfbox}{\textsc{PDFBox}\xspace}
\newcommand{\graphhopper}{\textsc{GraphHopper}\xspace}
\newcommand{\gephi}{\textsc{Gephi}\xspace}
\newcommand{\OO}{\textbf{OO}\xspace}
\newcommand{\PO}{\textbf{PO}\xspace}
\newcommand{\CO}{\textbf{CO}\xspace}

\title{Mimicking Production Behavior with \\Generated Mocks}

\author{
\IEEEauthorblockN{
Deepika Tiwari,
Martin Monperrus,
and Benoit Baudry}

\IEEEauthorblockA{
KTH Royal Institute of Technology, Stockholm, Sweden\\
\{deepikat, monperrus, baudry\}@kth.se}
}

\date{July 2022}

\begin{document}

\maketitle

\begin{abstract}
Mocking allows testing program units in isolation. 
A developer who writes tests with mocks faces two challenges: 
design realistic interactions between a unit and its environment; and understand the expected impact of these interactions on the behavior of the unit. 
In this paper, we propose to monitor an application in production to generate tests that mimic realistic execution scenarios through mocks.
Our approach operates in three phases. First, we instrument a set of target methods for which we want to generate tests, as well as the methods that they invoke, which we refer to as mockable method calls. Second, in production, we collect data about the context in which target methods are invoked, as well as the parameters and the returned value for each mockable method call. 
Third, offline, we analyze the production data to generate test cases with realistic inputs and mock interactions. The approach is automated and implemented in an open-source tool called \rick. 
We evaluate our approach with three real-world, open-source Java applications.
\rick  monitors the invocation of $128$ methods in production across the three applications and captures their behavior.
Based on this captured data, \rick generates test cases that include realistic initial states and test inputs, as well as mocks and stubs.
All the generated test cases are executable, and $52.4\%$ of them successfully mimic the complete execution context of the target methods observed in production. 
The mock-based oracles are also effective at detecting regressions within the target methods, complementing each other in their fault-finding ability.
We interview $5$ developers from the industry who confirm the relevance of using production observations to design mocks and stubs.  
Our experimental findings clearly demonstrate the feasibility and added value of generating mocks from production interactions.
\end{abstract}
\begin{IEEEkeywords}
Mocks, production monitoring, mock-based oracles, test generation
\end{IEEEkeywords}

\section{Introduction}
\IEEEPARstart{S}{oftware} testing is an expensive, yet indispensable, activity. It is done to verify that the system as a whole, as well as the individual modules that compose it, behave as expected. The latter is achieved through unit testing. When a unit interacts with others, or with external components, such as the file system, database, or the network, it becomes challenging to test it in isolation. As a solution to this problem, developers rely on a technique called mocking \cite{mackinnon2000endo}. Mocking allows a unit to be tested on its own, by replacing dependent objects that are irrelevant to its functionality, with ``fake`` implementations. There are several advantages of mocking, such as faster test executions, fewer side-effects, and quicker fault localization \cite{thomas2002mock}. 

Despite their advantages, using mocks within tests requires considerable engineering effort.
Developers must first identify the components that may be replaced with mocks \cite{spadini2017mock}.
They must also determine how these mocks behave when triggered with a certain input, i.e., how they are stubbed \cite{barr2014oracle, zhu2023stubcoder}. 
In order to address these challenges, several techniques have been developed to automatically generate mocks.
For example, mocks have been generated through search-based algorithms by including contextual information, such as the state of the environment, in the search space for input data generation \cite{arcuri2014automated, arcuri2015generating}.
Dynamic symbolic execution has been used to define the behavior of mocks through generated mock classes \cite{islam2010dsc+}.
System test executions can be monitored to derive unit tests that use mocks \cite{saff2005automatic}.
This preliminary research has validated the concept of mock generation.
However, the test inputs and the mocks produced by these techniques are either synthetic or manually written by developers per their assumptions of how the system behaves. 
These approaches do not guarantee that the generated mocks reflect realistic behaviors as observed in production contexts.

The fundamental premise of mocking is to replace a real object with a fake one that mimics it \cite{fowlerstubs, christakis2017general}.
This implies that, for it to be useful, the behavior of the mock should resemble, as closely as possible, that of a real object \cite{daka2014survey}. 
Our key insight is to derive realistic behavior from real behavior, i.e., to generate mocks from production usage.
This builds upon the fact that 1) the behavior of an application in production is the reference one \cite{wang2017behavioral}, and 2) this production behavior can be used for test generation  \cite{9240614, 9526340}.
Given a set of methods of interest for test generation, we monitor them in production.
As a result of this monitoring, we capture realistic execution contexts to generate tests for each target method, where external objects are replaced with mocks, and stubbed based on production states.

In this paper, we propose \href{https://youtu.be/dQw4w9WgXcQ}{\rick}, an approach for mimicking production behavior with generated mocks.
\rick monitors an application executing  in production with the goal of generating valuable tests.
The intention of the generated tests is to verify the behavior of the methods, where the reference behavior captured in the oracle is the one from production.
The interactions of this method with other external objects are mocked. 
Within each generated test, the data captured from production is expressed as rich serialized test inputs.
Each generated test has a mock-based oracle, which verifies distinct aspects of the invocation of the target method and its interactions with mock methods, such as the object returned from the target method, the parameters with which the mock methods are called, as well as the frequency and sequence of these mock method calls.

Our approach for the synthesis of mocks is based on two fundamentally novel concepts.
First, the mocks are stubbed with data and interactions observed in production. 
Second, our three mock-based oracles enable powerful behavior verification.
A key benefit of this latter point is that, in addition to checking the output of a method directly with a straightforward assertion, we also verify the actions that should occur within the  method.

We evaluate the capabilities of \rick with three open-source Java applications: a map-based routing application called \graphhopper, a feature-rich graph analysis and visualization tool called \gephi, and a utility library for working with PDF documents called \pdfbox. 
We target $212$ methods across the three applications, which get invoked as the applications execute in typical production scenarios.
\rick generates $294$ tests for $128$ of these methods.
Within each generated test, \rick recreates execution states that mimic production ones with objects that range in size from $37$ bytes to $39$ megabytes.
{When we execute the tests, we find that $68$ of the $128$ methods ($53.1$\%) have at least one passing test that recreates real usage conditions, and $154$ of the $294$ ($52.4$\%) tests successfully recreate the complete production execution context.} These results indicate that \rick is capable of monitoring applications in production, capturing realistic behavior for target methods, and transforming it into tests that mimic the behavior of the methods, while isolating it from its interactions with external objects. 
{Furthermore, through mutation analysis, we determine that the generated tests are effective at detecting regressions within the target methods.
The mock-based oracles contained in the generated tests complement each other with respect to their ability to detect bugs.}
To assess the quality of the \rick-generated tests, we interview $5$ software developers from different sectors of the IT industry. All of them find the collection of production values to be relevant to generate realistic mocks. Moreover, they appreciate the structure and the understandability of the tests generated  by \rick.

To sum up, the key contributions of this paper are:
\begin{itemize}
    \item The novel concept of the automated generation of mocks, stubs, and oracles using data collected from production.
    \item A comprehensive methodology for generating tests that mimic complex production interactions through mocks, by capturing receiving objects, parameters, returned values, and method invocations, for a method under test, while an application executes.
    \item An evaluation of the approach on $3$ widely-used, large, open-source Java applications, demonstrating the feasibility and benefits of generating mocks.
    \item A publicly available tool called \rick implementing the approach for Java, and a replication package for future research on this novel topic\footnote{\url{https://zenodo.org/doi/10.5281/zenodo.6914463}}.
\end{itemize}

The rest of the paper is organized as follows. \autoref{sec:background} presents the background on mocking and mock objects. We describe how \rick generates tests and mocks in \autoref{sec:rick}. Next, \autoref{sec:methodology} discusses the methodology we follow for our experiments, applying \rick to real-world Java projects, followed by the results of these experiments in \autoref{sec:results}.
\autoref{sec:related-work} presents closely related work, and \autoref{sec:conclusion} concludes the paper.

\section{Background}\label{sec:background}

This section summarizes the key concepts of mock objects, and how they are used in practice. We also discuss the challenges of using mocks within tests.

\subsection{Mock Objects}
Software comprises of individual modules or units. These units interact with each other, as well as with external libraries, for example, to send emails, transfer data over the network, or perform database operations. This facilitates modular development, as different teams can work in parallel on implementing distinct functionalities of the system. The modules are then composed together, in order to achieve use cases. Yet, a disadvantage of this coupling is that testing each unit in isolation from others is not straightforward.
\emph{Mocking} was proposed as a solution to this problem \cite{mackinnon2000endo}. It is a mechanism to replace real objects with skeletal implementations that mimic them \cite{thomas2002mock}.
Mocking allows individual functionalities to be tested independently. 
The process of unit testing with mocks is faster and more focused \cite{spadini2017mock}. Since the test intention is to verify the behavior of one individual unit, mocking can facilitate fault localization.
External objects, with interactions that are complex to set up within a test, can be replaced with mocks \cite{9411706}. Furthermore, a test can be made more reliable by using mocks to replace external, potentially unreliable or non-deterministic, components \cite{arcuri2014automated, arcuri2015generating}.
Mock objects typically behave in specific, pre-determined ways through a process called \emph{stubbing} \cite{fowlerstubs, zhu2023stubcoder}. For example a method called \texttt{getAnswer} invoked on a mock object can be stubbed to return a value of \texttt{42}, without the actual invocation of \texttt{getAnswer}. Stubbing can be very useful for inducing behavior that may be hard to produce locally within the test, such as error- or corner-cases.
Mocking can also be used for \emph{verifying} object protocols \cite{beckman2011empirical}. For example, consider a method called \texttt{subscribeToNewsletter}, which should call another method \texttt{sendWelcomeEmail} on an object of type \texttt{EmailService}. Developers can mock the \texttt{EmailService} object within the test for \texttt{subscribeToNewsletter}, to verify that the method \texttt{sendWelcomeEmail} is indeed invoked on \texttt{EmailService} exactly once, with a parameter of type \texttt{UserID}. This interaction is therefore verifiable without side-effects, i.e., without an actual email being sent.

In short, the three key concepts of testing with mocks are \emph{Mocking}, \emph{Stubbing}, and \emph{Verifying}. Real objects can be replaced with fake implementations called mocks. These mocks can be stubbed to define tailor-made behaviors, i.e., produce a certain output for a given input. Moreover, the interactions made with the mocks can be verified, such as the number of times they were triggered with a given input, and in a specific sequence.

\subsection{The Practice of Testing with Mocks}
Mocks can be implemented in several ways. For example, developers may manually write classes that are intended as replacements for real implementations \cite{9240675}. However, a more common way of defining and using mocks is through the use of mocking libraries, which are available for most programming languages. Mockito\footnote{\url{https://site.mockito.org/}} is one of the most popular mocking frameworks for Java, both in the industry and in software engineering research \cite{spadini2019mock, 6958396}. It can be integrated with testing frameworks such as JUnit and TestNG, allowing developers to write tests that use mocks.

\begin{lstlisting}[language=Java, belowskip={-10pt}, label={lst:bg-example-mut}, caption={Target method \texttt{purchaseTickets} has mockable method calls on the \texttt{PaymentService} object}, float]
public class ReservationCentre {
  ...
  // Target method
  public void purchaseTickets(int quantity, PaymentService paymentService) {
    ...
    double amount = basePrice * quantity;
    ...
    // Mockable method call #1
    if (paymentService.checkActiveConnections() > 0) {
      ...
      // Mockable method call #2
      boolean isPaymentSuccessful = paymentService.processPayment(amount);
      ...
    }
    ...
    return ...;
  }
  ...
}
\end{lstlisting}

\begin{lstlisting}[language=Java, belowskip={-10pt}, label={lst:bg-example-test}, caption={A test for the \texttt{purchaseTickets} method which mocks \texttt{PaymentService}}, float]
@Test
public void testTicketPurchasing() {
  ReservationCentre resCentre = new ReservationCentre();
  
  // Mock external types
  PaymentService mockPayService = mock(PaymentService.class);

  // Stub behavior
  when(mockPayService.checkActiveConnections()).thenReturn(1);
  when(mockPayService.processPayment(42.24)).thenReturn(true);

  resCentre.purchaseTickets(2, mockPayService);
  
  // Verify invocations on mocks
  verify(mockPayService, times(1)).checkActiveConnections();
  verify(mockPayService, times(1)).processPayment(anyDouble());
}
\end{lstlisting}

Let us consider the example of the method \texttt{purchaseTickets} presented in \autoref{lst:bg-example-mut}. This method handles the purchase of tickets, including the interaction with the payment gateway, \texttt{PaymentService}. It is defined in the \texttt{ReservationCentre} class, and takes two parameters. The first parameter is an integer value for the \texttt{quantity} of tickets, and the second parameter is the object \texttt{paymentService} of the external type \texttt{PaymentService}. Two methods are called on the \texttt{paymentService} object: \texttt{checkActiveConnections} on line $9$, and \texttt{processPayment} on line $12$ which accepts a parameter of type \texttt{double}. We illustrate the use of mocks, stubs, and verification through the unit test for \texttt{purchaseTickets} presented in \autoref{lst:bg-example-test}. The intention of this test, \texttt{testTicketPurchasing}, is to verify the behavior of \texttt{purchaseTickets}, while mocking its interactions with \texttt{PaymentService}. First, the receiving object \texttt{resCentre} of type \texttt{ReservationCentre} is set up (line $3$). Next, \texttt{PaymentService} is mocked, through the \texttt{mockPayService} object (line $6$). Lines $9$ and $10$ stub the two methods called on this mock: \texttt{checkActiveConnections} is stubbed to return a value of \texttt{1}, and \texttt{processPayment} is stubbed to return \texttt{true} when invoked with the \texttt{double} value \texttt{42.24}. Finally, on line $12$, \texttt{purchaseTickets} is called with the \texttt{quantity} of $2$, and the mocked parameter \texttt{mockPayService}. The statements on lines $15$ and $16$ verify that this invocation of \texttt{purchaseTickets} calls \texttt{checkActiveConnections} exactly once, and \texttt{processPayment} exactly once, with a double value (specified using \texttt{anyDouble()}). Thus, this test verifies the behavior of the target method, \texttt{purchaseTickets}, isolating it from the interactions with a real \texttt{PaymentService} object. Moreover, method calls on \texttt{PaymentService} are stubbed so that \texttt{purchaseTickets} gets executed as it normally would, without the side-effect of an actual payment being made. This allows for more focus on the method under test, \texttt{purchaseTickets}.

\subsection{The Challenges of Mocking}\label{sec:mocking-challenges}
Despite the benefits of mocking that we highlight, it is not trivial to incorporate mocks in practice. Deciding what to mock, and how the mocks should behave, is hard \cite{christakis2017general}. For example, developers would first identify that \texttt{PaymentService} may be mocked within the test for \texttt{purchaseTickets} in \autoref{lst:bg-example-test}. Next, they must also manually define concrete values for the parameters and returned values for stubbing the calls made on this mock, in order trigger a specific path through \texttt{purchaseTickets}. Additionally, they would have to determine which interactions made on this mock are verifiable.
It is also challenging to decide between conventional object-based testing and mock-based testing.
\revisedthree{
Because of these challenges, developers are hesitant to incorporate mocks within their testing practice, as highlighted by a study by Spadini \textit{et al.} \cite{spadini2019mock}, who found that mocks are most likely to be introduced at the time a test class is first written.
This suggests the potential opportunity and benefits of automated mock generation throughout the development lifecycle.
The results from our developer study in RQ5 address this aspect.}

To address the challenges of manually implementing mocks, several studies propose methodologies for their automated generation, such as through search-based algorithms \cite{ArcuriFJ17} or symbolic execution \cite{tillmann2006mock}. 
However, none of these studies use data from production executions to do so. In this work, we propose to monitor an application in production, with the goal of generating tests with mocks. These tests use mocks to 1) isolate a target method from external units, and 2) verify distinct aspects of the behavior with oracles specific to mocks. 

\section{Mock Generation with \rick}\label{sec:rick}

We introduce \rick, a novel approach for automatically generating tests with mock objects, using data collected during the execution of an application. 
In production, \rick collects realistic data for recreating the program states for the method under test, as well as the parameters and values returned by methods called on external objects.
In \autoref{sec:rick-overview}, we present an overview of the \rick pipeline. This is followed in \autoref{sec:rick-oracles} by a discussion of the kinds of oracles produced by \rick. Next, \autoref{sec:rick-design} motivates the design decisions of \rick and highlights its key features.
We discuss in \autoref{sec:rick-sdlc} how \rick can be useful in the software development lifecycle.
Finally, \autoref{sec:rick-implementation} presents technical details of its implementation.

\subsection{Overview}\label{sec:rick-overview}

\begin{figure*}
\centering
\includegraphics[width=\textwidth]{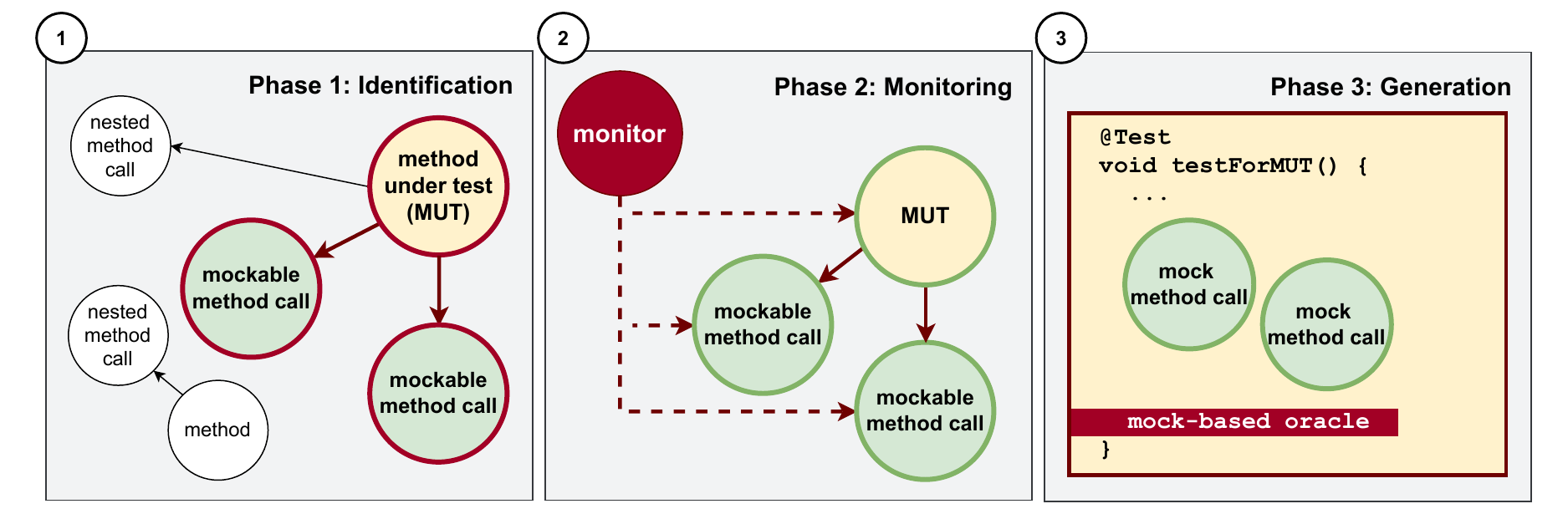}
\caption{The \rick test generation pipeline: offline, identify methods under test and mockable method calls; in production, monitor state and arguments for methods under test and mockable method calls; offline again, generate tests with mocks for the methods under test}
\label{fig:rick-overview}
\end{figure*}

\rick operates in three phases. We illustrate them in \autoref{fig:rick-overview}.
In the first phase \circled{1}, \rick identifies test generation targets within an application. These targets are called methods under test, and they have mockable method calls. We define them as follows:

\emph{\textbf{Methods under Test (MUTs)}:} The target methods for test generation by \rick are called methods under test (MUTs). A method is considered as being an MUT if it invokes methods on objects of other types. The identification of MUTs forms the basis of the test generation effort, since the intention of each test generated by \rick is to verify the behavior of one MUT after isolating it from such external interactions.

\emph{\textbf{Mockable method calls}:} We define a mockable method call as a method call nested within an MUT, that is made on a field or a parameter object whose type is different from the declaring type of the MUT. 
\revisedthree{\rick will mock objects of types that are declared within the project, and not types from the standard library or dependencies.}
When \rick generates a test for the MUT, a mockable method call becomes a \emph{mock method call}, i.e., the external object is replaced with a mock object, and the mockable method call occurs on this mock object.

\autoref{fig:rick-overview} illustrates the identification of an MUT (highlighted in yellow), together with its mockable method calls (shown here in green circles).
In \autoref{sec:rick-design} we detail how a nested method call qualifies as being mockable.
As an example, consider the class \texttt{ClassUnderTest} presented in \autoref{lst:target-methods}.
\rick identifies the method \texttt{methodUnderTestOne} (line $5$) as an MUT. Moreover, the nested call to \texttt{mockableMethodOne} on line $8$ within \texttt{methodUnderTestOne} is identified as a mockable method call because it is made on the field \texttt{extField} of \texttt{ClassUnderTest} (line $2$), which is of an external type. Similarly, the method \texttt{methodUnderTestTwo} is also considered as an MUT by \rick, because it has two mockable methods called within it. The first is \texttt{mockableMethodTwo} called $42$ times inside a loop (line $20$), and the second one is \texttt{mockableMethodThree} (line $22$). Both of these methods are called on \texttt{extParam}, which is an external parameter of \texttt{methodUnderTestTwo}.

\revisedthree{
The second phase \circled{2} of \rick occurs when the application is deployed and running.
During this phase, \rick monitors the invocation of the MUTs identified in the previous phase, and collects data corresponding to these invocations.
By construction, the monitoring data reflects real interactions by end-users.
Moreover, for inadequately-tested applications, it may represent usage scenarios that are not well tested by developer-written tests \cite{wang2017behavioral,9526340}.}

\rick collects data for an invocation in production with the end goal of recreating the same invocation within a generated test. This data includes the parameters and returned values for each MUT and its corresponding mockable method calls, as well as the object on which the MUT is invoked, which we refer to as the \emph{receiving object}. \autoref{fig:rick-overview} depicts the second phase, where the monitor defined within \rick is attached to both an MUT as well as its mockable method calls, in order to collect data about their invocations. \autoref{sec:rick-design} presents more details on this monitoring.

Finally, in the third phase \circled{3}, \rick uses the data collected in the second phase as inputs to generate tests with mocks, as illustrated in \autoref{fig:rick-overview}. These tests are designed to recreate the invocation of the MUTs, and verify their behavior as was observed in production, while simulating the interactions of the MUTs with external objects using mocks and stubs.
They can serve as regression tests, and can also potentially lead to faster fault localization because they isolate the invocation of the MUT from the mockable method calls. For example, \autoref{lst:test-oo} presents one test generated for the MUT \texttt{methodUnderTestOne}. This test verifies the observed behavior of \texttt{methodUnderTestOne} through the assertion on line $17$, while mocking the \texttt{ExternalTypeOne} object (line $8$). Within the test, the mockable method \texttt{mockableMethodOne} becomes the mock method, and is stubbed on line $11$ using its parameter and returned value captured from production.
We present more information on how \rick processes production data to generate tests in \autoref{sec:rick-design}.
The generated tests are the final output of the \rick pipeline. Each generated test falls under one of three categories, determined by the kind of oracle within it. We discuss these categories in \autoref{sec:rick-oracles}.

\begin{lstlisting}[language=Java, belowskip={-10pt}, label={lst:target-methods}, caption={The class, \texttt{ClassUnderTest}, has four methods. \rick identifies two of these methods, \texttt{methodUnderTestOne} and \texttt{methodUnderTestTwo}, as MUTs as they have mockable method calls.}, float]
class ClassUnderTest {
  ExternalTypeOne extField;
  ...
  
  public int methodUnderTestOne(int param) {
    ...
    // mockable method call on field
    int x = param + extField.mockableMethodOne(booleanVal);
    // nested, non-mockable method call
    int y = x + methodFour();
    ...
    return ...;
  }
  
  public int methodUnderTestTwo(double param,
                              ExternalTypeTwo extParam) {
    ...
    // mockable method calls on parameter
    for (int i = 0; i < 42; i++) {
      listOfInts.add(extParam.mockableMethodTwo(floatVals[i]));
    }
    int z = extParam.mockableMethodThree(intVal);
    ...
    return ...;
  }
  
  private void methodThree() { ... }
  
  public int methodFour() {
    // nested, non-mockable method call
    methodThree();
    ...
    return ...;
  }
}
\end{lstlisting}

\subsection{Mock-based Oracles}\label{sec:rick-oracles}
The oracle in a unit test specifies a behavior that is expected as a consequence of running the MUT with a specific test input \cite{barr2014oracle}.
In the context of the tests generated by \rick, the oracles in the generated test verify the behavior of the MUT, while isolating it from method calls to external objects made within the MUT, i.e., mockable method calls. This facilitates the decoupling of the MUT from its environment, and allows the focus of the testing to be on the MUT itself.
Moreover, the behavior being verified in the generated tests, both for the MUT and the mockable method calls, is sourced from production. This means that through these generated tests, developers can verify how the system behaves for actual users.

There is no systematized knowledge on oracles for tests with mocks. To overcome this, we now define three categories of oracles, all implemented by \rick.

\textbf{Output Oracle, OO}:
\revisedthree{The first category of tests generated by \rick have an output oracle. 
This oracle verifies that the behavior of the MUT is the same as the one observed in production, despite the introduction of mock objects.
Even though this oracle relies on regular assertions, we still consider it as a mock-based oracle, as it assesses the behavior of the MUT in the presence of mock objects.
}
A failure in a test with an output oracle indicates a regression in the MUT, which may  be caused by its interaction with the mockable method call.
\autoref{lst:test-oo} presents an example of a generated test with an output oracle. The MUT  is \texttt{methodUnderTestOne}, and the mockable method call is \texttt{mockableMethodOne}. This test corresponds to one invocation of \texttt{methodUnderTestOne} observed by \rick in production. The test recreates the receiving object, \texttt{productionObj}, as it was observed in production, by deserializing it (line $5$). Next, it mocks the field \texttt{extField} and injects it into \texttt{productionObj} (line $8$). This is followed in line $11$ by stubbing \texttt{mockableMethodOne}, to return a value of \texttt{27} when invoked with the parameter \texttt{true}, in accordance with the observed production behavior of \texttt{mockableMethodOne}. On line $14$, \texttt{methodUnderTestOne} is invoked on \texttt{productionObj} with the production parameter \texttt{17}. Finally, the output oracle is the assertion on line $17$, which verifies that the output from this invocation of \texttt{methodUnderTestOne}, with the stubbed call to \texttt{mockableMethodOne}, is \texttt{42}, which is the value observed for this invocation in production.
\begin{lstlisting}[language=Java, belowskip={-10pt}, label={lst:test-oo}, caption={A \rick test with an Output Oracle, \OO, for the MUT \texttt{methodUnderTestOne}}, float]
@Test
public void testMethodUnderTestOne_OO() {
// Arrange
  // Create test fixture from serialized production data
  ClassUnderTest productionObj = deserialize(new File( "receiving1.xml"));
  
  // Inject the mock
  ExternalTypeOne mockExternalTypeOne = injectMockField_extField_InClassUnderTest();
  
  // Stub the behavior
  when(mockExternalTypeOne.mockableMethodOne(true)) .thenReturn(27);
  
// Act
  int actual = productionObj.methodUnderTestOne(17);
  
// Assert
  assertEquals(42, actual);
}
\end{lstlisting}

\textbf{Parameter Oracle, PO}: 
The second category of generated tests have an oracle to verify that the mockable method calls occur with specific parameter(s), the same as production, within the invocation of the MUT.
A test with a parameter oracle may fail due to regressions in the MUT which cause a mockable method call to be made with unexpected parameters.
An example of this oracle is presented in \autoref{lst:test-po}. This test recreates the receiving object \texttt{productionObj}, for the MUT \texttt{methodUnderTestTwo} (line $5$). Next, it prepares a mock object for \texttt{ExternalTypeTwo} called \texttt{mockExternalTypeTwo} (line $8$), and stubs the $42$ invocations of the mockable method call, \texttt{mockableMethodTwo}. For brevity, we include only two of these stubs on lines $11$ and $12$. The single invocation of \texttt{mockableMethodThree} (line $14$) is also stubbed, with the parameter and returned value observed in production. This is followed by a call to \texttt{methodUnderTestTwo} on \texttt{productionObj} (line $17$), passing it \texttt{mockExternalTypeTwo} as parameter. Finally, the statements on lines $20$ to $23$ are unique to this category of tests, and serve as the parameter oracle. The statements on lines $21$ and $22$ verify that \texttt{mockableMethodTwo} is called at least once on \texttt{mockExternalTypeTwo} with the concrete production parameter \texttt{4.2F}, as well as with \texttt{9.8F}. We omit the verification of the other $40$ invocations of \texttt{mockExternalTypeTwo} from this code snippet. Next, the parameter oracle verifies on line $23$ that \texttt{mockableMethodThree} is called at least once with the parameter \texttt{15}, within this invocation of \texttt{methodUnderTestTwo}.
\begin{lstlisting}[language=Java, belowskip={-10pt}, label={lst:test-po}, caption={A \rick test with a Parameter Oracle, \PO, for the MUT \texttt{methodUnderTestTwo}}, float]
@Test
public void testMethodUnderTestTwo_PO() {
// Arrange
  // Create test fixture from serialized production data
  ClassUnderTest productionObj = deserialize(new File( "receiving2.xml"));
  
  // Create the mock
  ExternalTypeTwo mockExternalTypeTwo = mock( ExternalTypeTwo.class);
  
  // Stub the behavior
  when(mockExternalTypeTwo.mockableMethodTwo(4.2F)) .thenReturn(89);
  when(mockExternalTypeTwo.mockableMethodTwo(9.8F)) .thenReturn(92);
  ...
  when(mockExternalTypeTwo.mockableMethodThree(15)) .thenReturn(48);
  
// Act
  productionObj.methodUnderTestTwo(6.2, mockExternalTypeTwo);
  
// Assert
  verify(mockExternalTypeTwo, atLeastOnce()) .mockableMethodTwo(4.2F);
  verify(mockExternalTypeTwo, atLeastOnce()) .mockableMethodTwo(9.8F);
  ...
  verify(mockExternalTypeTwo, atLeastOnce()) .mockableMethodThree(15);
}
\end{lstlisting}

\textbf{Call Oracle, CO}:
Oracles in the generated tests for the third category verify the sequence and frequency of mockable method calls within the invocation of the MUT.
Any deviation from the expected sequence and frequency of mockable method calls within an MUT will cause a test with a call oracle to fail. This can be helpful for developers when localizing a regression related to object protocols within the MUT.
An example of this oracle is presented in \autoref{lst:test-co}. This test first recreates the receiving object, \texttt{productionObj}, for the MUT \texttt{methodUnderTestTwo} (line $5$), stubs the mockable method calls to \texttt{mockableMethodTwo} and \texttt{mockableMethodThree} (lines $11$ to $14$), and invokes \texttt{methodUnderTestTwo} with the mocked parameter (line $17$). Next, the call oracle in this test verifies the number of times the mockable method calls occur within this invocation of \texttt{methodUnderTestTwo}, as well as the order in which these calls occur. This is achieved with the order verifier defined on line $20$.
The statements on lines $21$ and $22$ verify that \texttt{mockableMethodTwo} is invoked exactly $42$ times with a float parameter, and that these invocations are followed by one call to \texttt{mockableMethodThree} with an integer parameter, as was observed in production.

\begin{lstlisting}[language=Java, belowskip={-10pt}, label={lst:test-co}, caption={A \rick test with a Call Oracle, \CO, for the MUT \texttt{methodUnderTestTwo}}, float]
@Test
public void testMethodUnderTestTwo_CO() {
// Arrange
  // Create test fixture from serialized production data
  ClassUnderTest productionObj = deserialize(new File( "receiving2.xml"));
  
  // Create the mock
  ExternalTypeTwo mockExternalTypeTwo = mock( ExternalTypeTwo.class);
  
  // Stub the behavior
  when(mockExternalTypeTwo.mockableMethodTwo(4.2F)) .thenReturn(89);
  when(mockExternalTypeTwo.mockableMethodTwo(9.8F)) .thenReturn(92);
  ...
  when(mockExternalTypeTwo.mockableMethodThree(15)) .thenReturn(48);
  
// Act
  productionObj.methodUnderTestTwo(6.2, mockExternalTypeTwo);
  
// Assert
  InOrder orderVerifier = inOrder(mockExternalTypeTwo);
  orderVerifier.verify(mockExternalTypeTwo, times(42)) .mockableMethodTwo(anyFloat());
  orderVerifier.verify(mockExternalTypeTwo, times(1)) .mockableMethodTwo(anyInt());
  
}
\end{lstlisting}

\subsection{Key Phases}\label{sec:rick-design}
As outlined in \autoref{sec:rick-overview}, \rick operates in three phases. We now discuss these three phases in more detail.

\subsubsection{Identification of Test Generation Targets}
It is not possible to generate test cases with mocks for all methods with nested method calls. For example, a static method invoked within another method is typically not mocked \cite{spadini2017mock}. Also, it is not feasible to replace an object created within the body of a method, and subsequently mock the interactions made with it.
Therefore, \rick includes a set of rules to determine the methods that can be valid targets for the generation of test cases and mocks. It is also possible for developers to provide an initial set of methods of interest, which \rick can consider. 

\paragraph{Identifying MUTs} First, \rick finds methods that are part of the API of the application, i.e., methods that are public, non-abstract, non-deprecated, and non-empty \cite{tillmann2008pex, thummalapenta2009mseqgen}. These criteria have also been used previously to generate differential unit tests for open-source Java projects \cite{9526340}. Of these methods, \rick selects as MUTs the ones that invoke other methods on objects of external types.

\paragraph{Identifying Mockable Method Calls}

\revisedthree{
Second, \rick identifies the nested method calls within each of the selected MUTs which could be mocked.
An MUT may have several nested method calls, not all of which are suitable for mocking.
For it to be mocked, a nested method call must be invoked on a parameter or a field, such that a mock can be injected to substitute it in the generated test.
Next, the declaring type of this parameter or field should be different from the type of the MUT, in order to represent an interaction of the MUT with an external type, per the theory of mocking external resources.  
\rick stubs methods that return a primitive or \texttt{String} value.
Mocks are never returned from stubbed methods.
Nested method calls that meet all these criteria, are marked as a \emph{mockable method calls}.}

We illustrate the rules for target selection with the help of the excerpt of the class \texttt{ClassUnderTest} in \autoref{lst:target-methods}.
This excerpt includes a field as well as four methods defined in \texttt{ClassUnderTest}.
The first method, \texttt{methodUnderTestOne} (lines $5$ to $13$), accepts an integer parameter, and returns an integer value. The body of \texttt{methodUnderTestOne} includes a call to the method \texttt{mockableMethodOne(boolean)}, on the field \texttt{extField} (line $8$). This field is declared as being of type \texttt{ExternalTypeOne} in \texttt{ClassUnderTest} (line $2$). There is another call on line $10$ to a method defined in \texttt{ClassUnderTest} called \texttt{methodFour}.
The second method, \texttt{methodUnderTestTwo} (lines $15$ to $25$), returns an integer value, and accepts two parameters. The first parameter is a double, and the second parameter called \texttt{extParam} is of type \texttt{ExternalTypeTwo}. Within the loop on lines $19$ to $21$, \texttt{methodUnderTestTwo} calls the method \texttt{mockableMethodTwo(float)} on the parameter \texttt{extParam} (line $20$). There is another call on \texttt{extParam} to \texttt{mockableMethodThree(int)} (line $22$).
The third method defined in \texttt{ClassUnderTest} is a private method called \texttt{methodThree} (line $27$). It does not call any other method.
Finally, the fourth method in this excerpt is \texttt{methodFour} (lines $29$ to $34$), which has a call to \texttt{methodThree} (line $31$).

As a consequence of the aforementioned criteria, \rick identifies \texttt{methodUnderTestOne} and \texttt{methodUnderTestTwo} as MUTs. Moreover, the nested method calls, \texttt{mockableMethodOne}, and \texttt{mockableMethodTwo} and \texttt{mockableMethodThree}, in these MUTs respectively, are recognized as mockable method calls by \rick. However, the call to \texttt{methodFour} within \texttt{methodUnderTestOne} is not mockable within \texttt{methodUnderTestOne}, and \texttt{methodThree} and \texttt{methodFour} are not MUTs since they do not fulfill these criteria.

\subsubsection{Monitoring Test Generation Targets}\label{sec:rick-monitoring}



\begin{figure}
\centering
\includegraphics[width=\columnwidth]{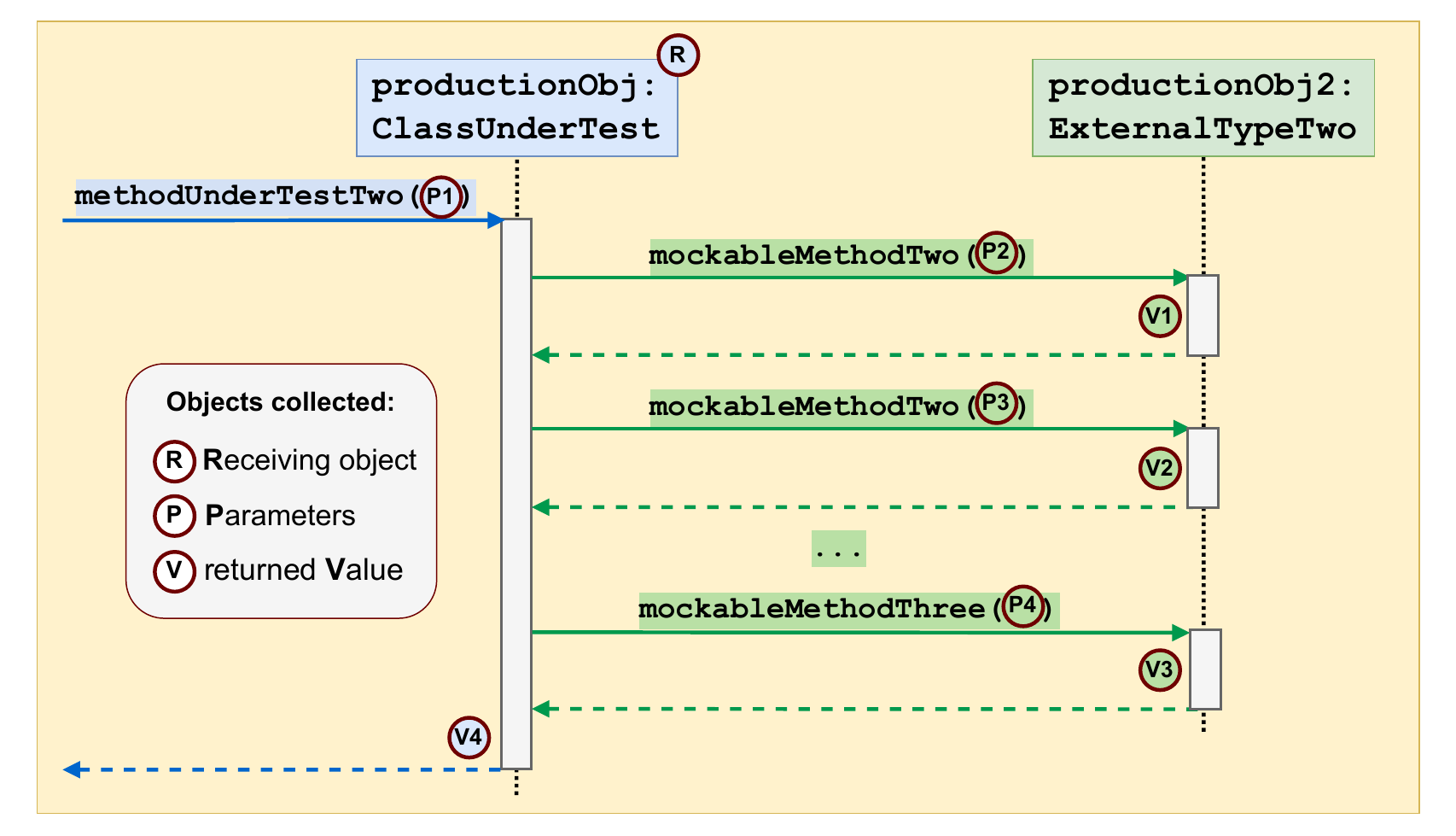}
\caption{Monitoring method invocations in production: \rick observes the invocation of the target method, \texttt{methodUnderTestTwo}, and captures the receiving object and parameters for this invocation, as well as the object returned from it. \rick also observes the method calls to \texttt{mockableMethodTwo} and \texttt{mockableMethodThree}, collecting their parameters and returned values.}
\label{fig:monitoring}
\end{figure}

Once it finds a set of MUTs and their corresponding mockable method calls, \rick instruments them in order to monitor their execution as the application runs in production.
The goal of this instrumentation is to collect data as the application executes.
\rick collects data about each invocation of an MUT: the receiving object, which is the object on which it is invoked, the parameters passed to the invocation, as well as the object returned from the invocation.
At the same time, \rick collects data about the mockable method calls within this MUT. This includes the parameters and the returned value for each mockable method call.
The data collected from this monitoring is serialized and saved to disk. 
For example, the sequence diagram in \autoref{fig:monitoring} illustrates the monitoring of the MUT \texttt{methodUnderTestTwo} in \texttt{ClassUnderTest} presented in \autoref{lst:target-methods}. For one invocation of \texttt{methodUnderTestTwo}, \rick collects its receiving object, parameters, and returned value, as well as the parameters and returned values corresponding to the invocations of mockable method calls to \texttt{mockableMethodTwo} and \texttt{mockableMethodThree} within \texttt{methodUnderTestTwo}. 

We systematically consider special cases.
First, a mockable method may be invoked from different MUTs. Also, an MUT may itself be a mockable method for another MUT. Moreover, an MUT may be invoked without its corresponding mockable method call(s), if the latter is invoked within a branch, for example. It is therefore important to ensure that the data collected for a mockable method call is associated with a specific invocation of an MUT. \rick implements this association by assigning a unique identifier to each MUT invocation, and the same identifier to each mockable method call within it. This information is required for the generation of all of the three kinds of oracles, i.e., the output oracle, the parameter oracle, as well as the call oracle.
Second, one invocation of an MUT may have multiple mockable method calls, which may or may not have the same signature. Furthermore, these invocations occur in a specific order within the MUT. In order to account for this, \rick collects the timestamps for each mockable method call. This is done to synthesize statements corresponding to the call oracle, which verify the sequence and frequency of mockable method calls in the generated tests.

\subsubsection{Generation of Tests with Mocks}
Once \rick has collected data about invocations of MUTs and corresponding mockable method calls, the final phase can begin, triggered by the developer. \rick connects an MUT invocation with mockable method calls by utilizing the unique identifiers assigned to each invocation observed in production. It then generates code to produce the three categories of tests for each invocation, as detailed in \autoref{sec:rick-oracles}. The final output from the test generation phase is a set of test classes, which include tests from the three categories, for each invocation of an MUT that was observed by \rick in production.

\rick generates tests by bringing together all the data it has observed, collected, and linked to the respective invocation of an MUT and its mockable method calls. Within each test generated by \rick for an MUT:
\begin{itemize}
    \item The serialized receiving object and parameter(s) for the MUT are deserialized to recreate their production state. For example, the receiving object of the respective MUT is recreated from its serialized XML state on line $5$ of \autoref{lst:test-oo}, \autoref{lst:test-po}, and \autoref{lst:test-co}.
    \item External objects, on which mockable method calls occur, are substituted with mock objects. For example, a mock object substitutes the external field \texttt{extField} on line $8$ of \autoref{lst:test-oo}. A mock object is prepared for the parameter \texttt{ExternalTypeTwo} on line $8$ of \autoref{lst:test-po} and \autoref{lst:test-co}.
    \item Mockable method calls become mock method calls: they are stubbed, with production parameter(s) and returned value. The mock method call within \texttt{methodUnderTestOne} is stubbed on line $11$ of \autoref{lst:test-oo}. The mock method calls within \texttt{methodUnderTestTwo} are stubbed from lines $11$ to $14$ in \autoref{lst:test-po} and \autoref{lst:test-co}.
    \item \revisedthree{The generated test case calls the method under test, once. This makes the test intention very clear: the behavior of the MUT is the one which will be assessed by the oracle. Note that multiple methods may be called on the mock objects, all stubbed. The number of mock objects and stubbed methods is what creates a large testing space.}    
    \item The oracle verifies a unique aspect about the invocation of the MUT and its interactions with the external object(s): the \OO on line $17$ of \autoref{lst:test-oo} verifies the output of \texttt{methodUnderTestOne}, the \PO on lines $20$ to $23$ of \autoref{lst:test-po} verify that the mock method calls occur with the same parameters as they did in production, and the \CO on lines $21$ and $22$ of \autoref{lst:test-co} verify that the mock method calls happen in the same sequence and the same frequency as they did in production.
\end{itemize}

\subsection{\rick in the Software Development Lifecycle}\label{sec:rick-sdlc}
\revisedthree{There are two main use cases for \rick in the software development pipeline.
First, for a project that has few automated tests, or uses only manual testing \cite{haas2021can,zetterlund2022harvesting}, \rick can be used to bootstrap the creation of a test suite.
The setup would be as follows: human QA testers are employed to evaluate and manually test the system as a whole.
Meanwhile, \rick would capture the realistic interactions that the testers trigger, and would generate automated unit test cases, which can be frequently run when the developers evolve the application.
}

\revisedthree{Second, for projects that already contain automated tests, \rick can contribute with unit tests that reflect realistic behavior, as observed in production.
Field executions can be a rich source of data, and are likely to include usage scenarios not envisioned by developers \cite{wang2017behavioral}.
Leveraging the monitoring and automated test generation capabilities of \rick would allow these behaviors to be incorporated into the test suite.
New tests based on production observations have been shown to complement developer-written tests and improve the effectiveness of the test suite \cite{9526340,alshahwan2024observation}.
}

\revisedthree{A key phase in both use cases is the curation of a set of essential methods of interest, which will be the targets for test generation with \rick.
Though \rick ships with good default filters for identifying target methods, developers can use their domain expertise to define the most valuable target methods within their project.
Ideal candidates for test generation include methods that have recently been added or modified, methods at the public interface of the application, or methods that do not meet a specified test adequacy criterion.
}

\revisedthree{The test cases generated by \rick fully depend on the production usages that were monitored.
To that extent, \rick is not good at generating tests for corner cases that rarely occur.
On the other hand, \rick is excellent at generating tests for mission-critical functionalities that reflect typical usage scenarios for a target application.
}

\revisedthree{Finally, the tests generated by \rick can be fully integrated in a code review process.
We envision that the lead test engineer handle the test generation and open a pull-request to add the new tests.
Then, fellow developers would review the tests generated for each target method before merging them into the test suite for the project.
This process can be repeated multiple times, one target method at a time, throughout the lifecycle of the project.
}

\subsection{Implementation}\label{sec:rick-implementation}
\rick is implemented in Java.
MUTs and their corresponding mockable method calls are identified through static analysis with Spoon \cite{pawlak2016spoon}.
Once identified, they are instrumented and monitored in production with Glowroot, an open-source Application Production Monitoring agent\footnote{\url{https://glowroot.org/}}. {Glowroot is a well-documented, industry-grade tool. It has a low overhead and is stable. This makes it the best fit for monitoring production executions for mock generation.}
\revisedthree{Data collection from production is handled through serialization by XStream\footnote{\url{https://x-stream.github.io/}}.}
\rick relies on the code generation capabilities of Spoon\footnote{\url{https://spoon.gforge.inria.fr/}} to produce JUnit tests\footnote{\url{https://junit.org/}}.
These tests define and use Mockito\footnote{\url{https://site.mockito.org/}} mocks, \revisedfour{specifically the \texttt{mockito-inline} flavour, which allows mocking final classes.}
\revisedfour{By default, \rick generates three separate tests that contain the parameter oracle the call oracle, and the output oracle. If the MUT does not return a primitive or a \texttt{String}, \rick generates only two tests with the parameter and the call oracles.}
\rick uses some capabilities provided by the \pankti framework \cite{9526340}.

\section{Experimental Methodology}\label{sec:methodology}
This section introduces our experimental methodology. We  describe the open-source projects we use as case studies to evaluate mock generation with \rick. Then, we describe the production conditions that we use to collect data for test generation. Next, we present our research questions and define the protocol that we use to answer them.

\subsection{Case Studies}\label{sec:case-studies}
As detailed in \autoref{sec:rick}, \rick uses data collected from an application in production, in order to generate tests with mock-based oracles.
For this evaluation, we therefore target applications that we can build, deploy, and for which we can define a production-grade usage scenario.
We manually search for three notable, open-source Java projects that satisfy these criteria. 
\revisedthree{We also make sure that the case studies represent different categories of software: a library with a command-line interface, a desktop application, and a backend server application.}

\revisedthree{\autoref{tab:case-studies} summarizes the details of the projects we use as case studies to evaluate the capabilities of \rick.
For each case study, we provide the exact version as well as the SHA of the commit used for our experiments. This information will facilitate further replication. 
We also provide the number of lines of code, the number of commits, and the number of methods as indicators of the scale of the project, as well as the number of stars in the project repository, as an indicator of its visibility.}
The last row in \autoref{tab:case-studies} indicates the number of candidate MUTs for each case study, which is the set of MUTs with mockable method calls identified by \rick.

\begin{table}
\renewcommand*{\arraystretch}{2}
\centering
\caption{Case studies for the evaluation of \rick}\label{tab:case-studies}
\begin{tabular}{l|
>{\columncolor{blue!5}}r|
>{\columncolor{green!5}}r|
>{\columncolor{yellow!5}}r}
\hline
\textbf{\textsc{Metric}} & \textbf{\graphhopper} & \textbf{\gephi} & \textbf{\pdfbox}\\
\hline
\textsc{Version} & 5.3 & 0.9.6 & 2.0.24 \\ \hline
\textsc{sha} &  
\texttt{\href{https://github.com/graphhopper/graphhopper/tree/af5ac0b0ae024da4b23966b1eef480508b5eedbb}{af5ac0b}} &

\texttt{\href{https://github.com/gephi/gephi/tree/ea3b28f3c670c3ba0762aed8f62694ba929e6ca5}{ea3b28f}} &

\texttt{\href{https://github.com/apache/pdfbox/tree/8876e8e1a0adbf619cef4638cc3cea073e3ca484}{8876e8e}}

\\ \hline
\textsc{Stars} & 3.7K & 4.9K & 1.7K \\ \hline
\textsc{Commits} & 5.8K & 6.5K & 8.7K \\ \hline
\textsc{loc} & 89K & 35K & 165K \\ \hline
\textsc{Methods} & 4,104 & 2,117 & 9,102 \\ \hline
\textsc{Candidate\_Muts} & 356 & 115 & 319 \\ \hline
\end{tabular}
\end{table}

Our first case study is the web-based routing application based on OpenStreetMap called \graphhopper\footnote{\url{https://www.graphhopper.com/open-source/}}. It allows users to find the route between locations on a map, considering diverse means of transport and other routing information such as elevation.
\revisedthree{We use version 5.3 of \graphhopper, with $89$K lines of code (LOC), $5,844$ commits, and over $4$K methods.}
The project's repository on GitHub has $3.7$K stars.

The second case study is \gephi, an application for working with graph data\footnote{\url{https://gephi.org/}}. With $4.9$K stargazers on GitHub, \gephi is very popular, and has been adopted by both the industry and by researchers \cite{bastian2009gephi}. It allows users to import graph files, manipulate them, and export them in different file formats.
We use version $0.9.6$ of \gephi, which includes  $6,548$ commits and $126$K LOC.
For our evaluation, we exclude the GUI modules, as the generation of GUI tests has its own challenges \cite{memon2013automated} that are outside the scope of \rick.
\revisedthree{The $8$ modules of \gephi we consider are implemented in $35$K LOC and contain $2,117$ methods in total.}

The last case study is \pdfbox, a PDF manipulation command-line tool developed and maintained by the Apache Software Foundation\footnote{\url{https://pdfbox.apache.org/}} \cite{xiao4100265empirical}. It can extract text and images from PDF documents, convert between text files and PDF documents, encrypt and decrypt, and split and merge  PDF documents. 
\revisedthree{As highlighted in \autoref{tab:case-studies}, we use version 2.0.24 of \pdfbox, which has $165$K LOC, over $9$K methods, $8,797$ commits, and has been starred by $1.7$K GitHub users.}

To generate tests with \rick, the first step consists of identifying candidate MUTs, which \rick will instrument so their invocations can be monitored as the project executes in production.
According to the criteria introduced in \autoref{sec:rick-design}, \rick identifies and instruments a total of $790$ CANDIDATE\_MUTs across the three applications: $356$ in \graphhopper, $115$ in \gephi, and $319$ in \pdfbox. These methods have interactions with objects of external types, where these objects are either the parameters of the MUT, or a field defined within the declaring type of the MUT.

\subsection{Production Usage}\label{sec:production-usage}

\begin{table}
\renewcommand*{\arraystretch}{2}
\centering
\caption{\revisedfour{Characteristics of the workloads for the case studies in production: The four metrics are defined in \autoref{sec:production-usage}.}}\label{tab:production-usage}
\begin{tabular}{c|
>{\columncolor{blue!5}}r|
>{\columncolor{green!5}}r|
>{\columncolor{yellow!5}}r|
r}
\hline
\textbf{\textsc{Metric}} & \textbf{\graphhopper} & \textbf{\gephi} & \textbf{\pdfbox} & \textbf{\textsc{Total}}\\
\hline
{\begin{tabular}[c]{@{}c@{}}\textsc{Mut\_} \\ \textsc{Invoked} \end{tabular}} & 72 & 68 & 72 & 212 \\ \hline
{\begin{tabular}[c]{@{}c@{}}\textsc{Mockable\_} \\ \textsc{Invoked} \end{tabular}} & 81 & 63 & 55 & 199 \\ \hline
{\begin{tabular}[c]{@{}c@{}}\textsc{Mut\_} \\ \textsc{Invocations} \end{tabular}} & 73,025 & 21,548 & 7,429,800 & 7,524,525 \\ \hline
{\begin{tabular}[c]{@{}c@{}}\textsc{Mockable\_} \\ \textsc{Invocations} \end{tabular}} & 246,822 & 202,630 & 5,144,790 & 5,594,242 \\ \hline
\end{tabular}
\end{table}

\revisedthree{Once the candidate MUTs for an application are identified, the instrumented application is deployed and run under a certain workload. 
As \rick aims at consolidating test suites for mission-critical functionalities that everybody relies on, we design workloads that exercise common features.
We manually analyze the case studies' codebase and refer to their documentation, in order to design workloads that capture commonly used features and operations.
}

\autoref{tab:production-usage} summarizes the key characteristics of the workloads. Rows 2 and 3 capture the scope for test generation that we consider for our experiments. The number of candidate MUTs actually invoked in production is indicated by MUT\_INVOKED, while MOCKABLE\_INVOKED represents the number of distinct mockable methods invoked within these invoked MUTs. We also report the number of times the MUTs and their mockable methods are invoked in  the rows MUT\_INVOCATIONS and MOCKABLE\_INVOCATIONS, respectively.
The number of invocations of MUTs and mockable methods demonstrate the relevance and comprehensiveness of the production scenarios we design. They represent the extent to which we exercise the three applications in production, and reflect actual usage of their functionalities. 
\revisedthree{In total, \rick focuses on 212 target methods, that invoke 199 mockable method calls.
Our experiments trigger more than 7.5 million invocations of these 212 methods.}

\begin{figure}
\centering
\includegraphics[width=\columnwidth]{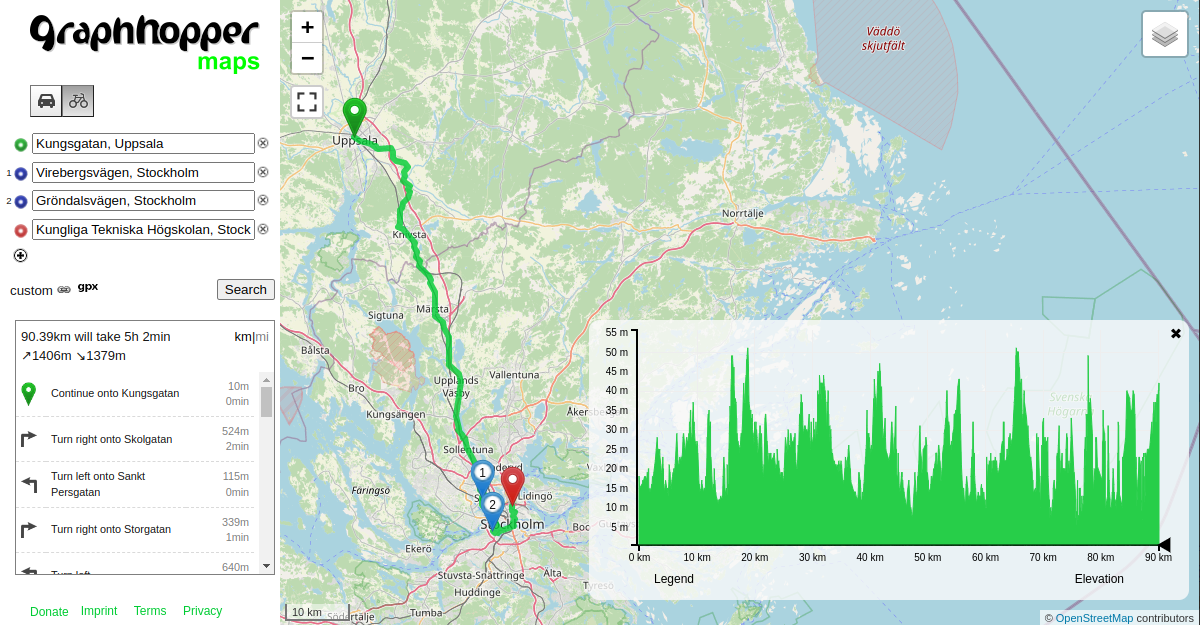}
\caption{Snapshot of \graphhopper in production. We query for the route between $4$ locations in Sweden, as \rick monitors target method invocations.}
\label{fig:graphhopper-prod}
\end{figure}

\textbf{\graphhopper:} To experiment with \graphhopper, we deploy its server and use its website to search for the car and bike route between four points in Sweden, specifically, from the residence of each author to their common workplace in Stockholm\footnote{\url{https://bit.ly/3LG2zSQ}}. \autoref{fig:graphhopper-prod} is a snapshot of this experiment. Recall from \autoref{sec:case-studies} that \rick monitors the invocation of $356$ CANDIDATE\_MUTs as \graphhopper executes. As presented in \autoref{tab:production-usage}, $72$ of the candidate MUTs are invoked (MUT\_INVOKED), which become test generation targets for \rick. Within these MUTs, $81$ distinct mockable methods are also called. With this production scenario, the $72$ MUTs are invoked a total of $73,025$ times, while the $81$ mockable methods are invoked $246,822$ times within the MUTs. 

\begin{figure}
\centering
\includegraphics[width=\columnwidth]{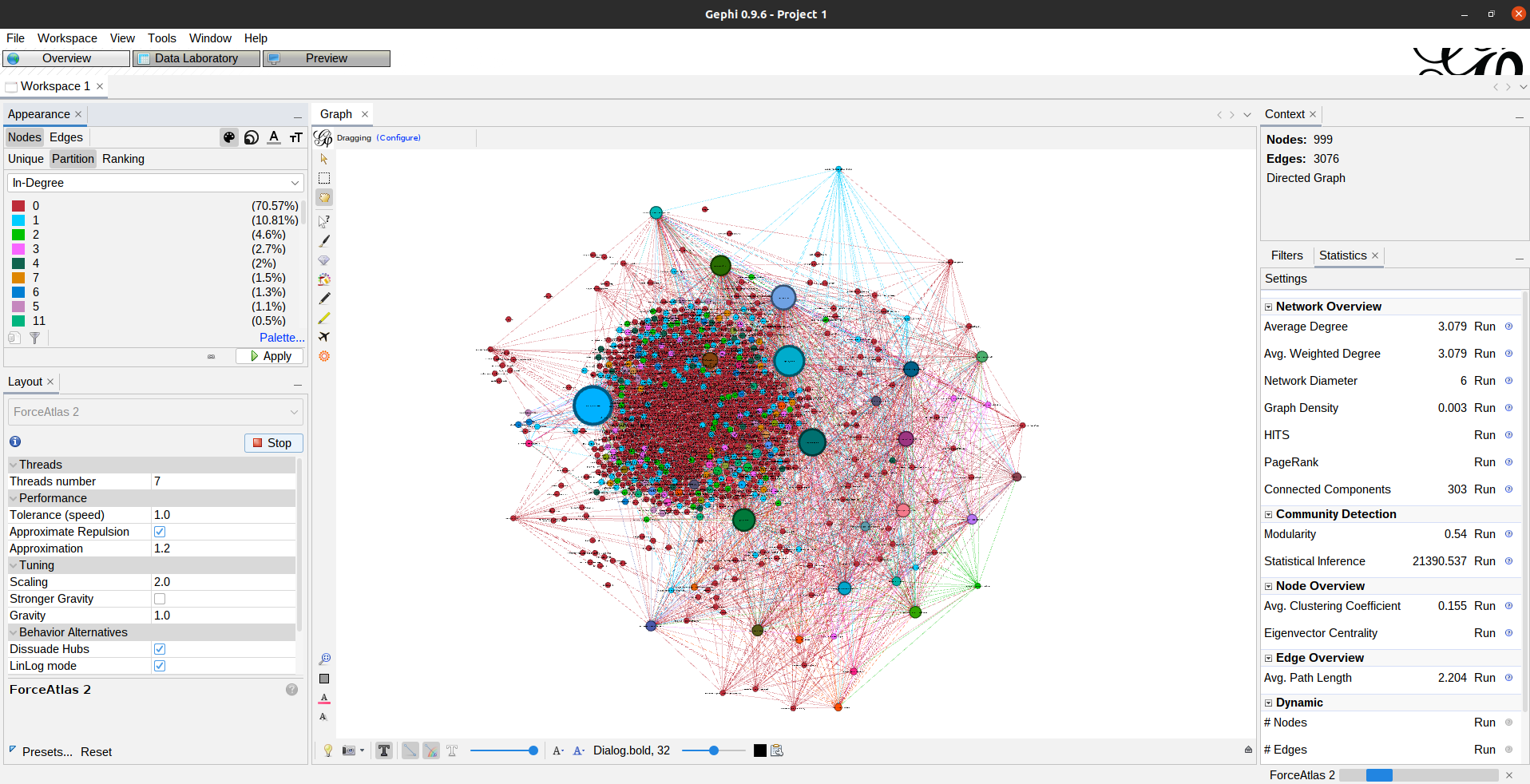}
\caption{We use \gephi to import a data file with details on $999$ Java artifacts on Maven Central. We interact with the features of \gephi to manipulate the resulting graph. \rick monitors method invocations corresponding to these interactions in production.}
\label{fig:gephi-prod}
\end{figure}

\textbf{\gephi:} As production usage for \gephi, we deploy the application and import a graph data file. This file has details about the top $999$ Java artifacts published on Maven Central, as well as the dependencies between them. We retrieve this data file from previous work \cite{Benelallam2019,Soto2019}. We use \gephi to produce a graph from this data, and to manipulate its layout, as illustrated in \autoref{fig:gephi-prod}. Finally, we export the resulting graph in PDF, PNG, and SVG formats, before exiting the application.
These interactions with \gephi lead to the invocation of $68$ of the $115$ candidate MUTs, and $63$ distinct mockable methods called by these MUTs. Moreover, these MUTs are invoked $21,548$ times, while there are $202,630$ mockable method calls.

\textbf{\pdfbox:}  We use the command-line utilities provided by \pdfbox to perform $10$ typical PDF manipulation operations on $5$ PDF documents. These documents are sourced from \cite{garfinkel2009bringing}, and the operations performed on them include text and image extraction, conversion into a text file, and vice-versa, etc. This methodology has been adopted from previous work \cite{9526340}.
Of the $319$ candidate MUTs, this workload  leads to the invocation of $72$ MUTs and $55$ different mockable methods. The MUTs are called over 7 million times, and the mockable methods are called over 5 million times. The magnitude of these invocation counts is due to the processing of a myriad of media content from the real-world PDF documents we select.

\subsection{Research Questions}\label{sec:rqs}
For each application described in \autoref{sec:case-studies} and exercised in production per the workload specified in \autoref{sec:production-usage}, \rick generates tests with mocks using the captured production data. Through these experiments, we aim to answer the following research questions.
\begin{itemize}
    \item \textbf{\emph{RQ1 [methods under test]}}: To what extent can \rick generate tests for MUTs invoked in production?
    \item \textbf{\emph{RQ2 [production-based mocks]}}: How rich is the production context reflected in the tests, mocks, and oracles generated by \rick?
    \item \textbf{\emph{RQ3 [mimicking production]}}: To what extent can the execution of generated tests and mocks mimic realistic production behavior?
    \item \textbf{\emph{RQ4
    [effectiveness]}}: How effective are the generated tests at detecting regressions?
    \item \textbf{\emph{RQ5 [quality]}}: What is the opinion of developers about the tests generated by \rick?
\end{itemize}

Each of the research questions presented above highlights a unique aspect of the capabilities of \rick, with respect to the automated generation of tests with mocks, using data sourced from production executions.

\subsection{Protocols for Evaluation}\label{sec:protocol}
The MUTs instrumented by \rick are invoked thousands of times (\autoref{tab:production-usage}). For experimental purposes, and to allow for a thorough, qualitative analysis of the results, we collect data about the first invocation of the MUT in production and use it to generate tests with \rick. We analyze these generated tests according to the following protocols in order to answer the research questions presented in \autoref{sec:rqs}.

\textbf{\emph{Protocol for RQ1}}: 
This first research question aims to characterize  the target MUTs for which \rick  transforms observations made in production into concrete test cases.
We describe the MUTs by reporting their number of lines of code (LOC) as well as the number of parameters.
We also report the number of tests generated by \rick for these target MUTs. This includes the tests with the three kind of mock-based oracles, \OO for MUTs that return primitive values, \PO, and \CO, for one invocation of the MUT observed  in production.

\textbf{\emph{Protocol for RQ2}}: 
With RQ2, we analyse the ability of \rick to
capture rich production contexts and turn them into test inputs  and oracles  that verify distinct aspects of their behavior.
We answer RQ2 by dissecting the data captured by \rick as the three applications execute, as well as the tests generated by \rick using this data.
First, to characterize the receiving object and parameters captured for the MUTs from production, we discuss the size of the serialized production state on disk. 
Second, we analyse the three kinds of oracles in the generated tests, specifically the assertion statement in the OO tests, and the number of verification statements in PO and CO tests. 
Furthermore, we report the number of external objects (fields and/or parameters) mocked within the tests, the stubs produced by \rick based on production observations, as well as the mock method calls.

\textbf{\emph{Protocol for RQ3}}: 
With RQ3 we highlight the feasibility and complexity of automatically generating tests and mocks that successfully execute in order to mimic  actual production behavior, in an isolated manner.
In order to answer RQ3, we execute the generated tests, and we analyse the outcome.
There are three possible outcomes of the execution of a generated test.
First, a test is successfully executed if the oracles  pass, implying that the test mimics the behavior of both the MUT and the mock method calls(s) observed by \rick in production.
Second, a generated oracle may fail, meaning that the test and its mocks do not replicate the production observations. For example, objects recreated within the generated test through deserialization may not be identical to those observed in production \cite{9526340}.
\revisedthree{Third, during the execution of the test, a runtime exception may happen before the oracles are evaluated, rendering them useless. We report on those cases as well.}

\textbf{\emph{Protocol for RQ4}}:
The goal of RQ4 is to determine how effectively the tests with mock-based oracles generated by \rick can detect regressions within the MUTs.
In order to do so, we inject realistic bugs within each MUT that has at least one passing \rick test \cite{just2014mutants}.
We rely on the LittleDarwin mutation testing tool \cite{parsai2017littledarwin} to generate a set of first-order mutants for these MUTs.
LittleDarwin is ideal for our analysis as it generates mutated source files on disk, which allows for automated, configurable, and reproducible experiments.
The $14$ mutation operators provided by LittleDarwin include the standard arithmetic, relational, and nullifying mutants, as well as one extreme mutation operator that replaces the whole body of an MUT with a default return statement \cite{vera2019comprehensive}.
Next, we substitute each MUT with a version that contains a mutant reachable by the test input \cite{petrovic2021practical}, and run each test generated for the MUT by \rick.
A test failure indicates that a mutant was covered, detected, and killed by the corresponding \OO, \PO, or \CO within the test.
We also analyze whether mock-based oracles differ in their ability to find faults \cite{StaatsGH12, GaySWH15a}.

\textbf{\emph{Protocol for RQ5}}: 
RQ5 is a qualitative assessment of the tests generated by \rick. It serves as a proxy for the readiness of the \rick tests to be integrated into the test suite of projects.
{To assess the quality of the generated tests, we conduct a developer survey, presenting a set of $6$ tests generated by \rick for $2$ MUTs in \graphhopper, to $5$ software testers from the industry.} 
\revisedfour{We carefully select the tests to present to the survey participants in order to have a representation of the three kinds of oracles as well as diverse mocking contexts i.e., external field or parameter objects.}
Developer surveys have previously been conducted to assess mocking practices \cite{spadini2017mock, daka2014survey}. The key novelty of our survey consists in assessing mocks that have been automatically generated.

We conduct each survey online for one hour, and follow this systematic structure: introduce mocking, the \rick test generation pipeline, and the \graphhopper case study; next, we give the participant access to a fork of \graphhopper on GitHub, with the \rick tests added, inviting  the participant to clone this repository, or browse through it online; finally, we ask them questions about the generated tests. 
{We select \graphhopper for the survey because its workload, fetching a route on a map, is intuitive and does not add to the complexity of the interview.
We select the MUTs for our discussions based on the following criteria: the MUTs have at least $10$ lines of code, and the tests generated for them have at least one stub.
From these, we select two MUTs in \graphhopper for which \rick has generated all three mock-based oracles, and for which the tests contain a mocked field and a mocked parameter.}

The goal of this survey is to gauge the opinion of developers about the quality of the $6$ tests with respect to three criteria: mocking effectiveness, structure, and understandability. Our replication package includes all the details about this survey.


\section{Experimental Results}\label{sec:results}
This section presents the results from our evaluation of \rick with \graphhopper, \gephi, and \pdfbox.
In \autoref{sec:results-rq1}, \autoref{sec:results-rq2} and \autoref{sec:results-rq3}, we answer RQ1, RQ2 and RQ3 based on the metrics summarized in 
\autoref{tab:graphhopper-results}, \autoref{tab:gephi-results}, and \autoref{tab:pdfbox-results}.
The results for RQ4 are presented in \autoref{sec:results-rq4}.
In \autoref{sec:results-rq5} we answer RQ5 based on the surveys conducted with testers from the industry.




\begin{table*}
\renewcommand*{\arraystretch}{1.6}
\centering
\caption{Experimental results for \graphhopper}\label{tab:graphhopper-results}
\resizebox{\textwidth}{!}{
\begin{tabular}{r|l|l|l?l|l|l|l|l|l|l?l|l|l}
\hline
\rowcolor{blue!5}
\multicolumn{4}{c?}{\textbf{\textsc{RQ1: Method Under Test}}} &
\multicolumn{7}{c?}{\textbf{\textsc{RQ2: Production-based Mocks}}} &
\multicolumn{3}{c}{\textbf{\textsc{RQ3: Mimicking Production}}} \\ \hline
\textbf{MUT\_ID} & 
\textbf{\#LOC} & 
\textbf{\#PARAMS} & 
\textbf{\#TESTS} & 
{\begin{tabular}[c]{@{}c@{}}\textbf{CAPTURED\_} \\ \textbf{OBJ\_SIZE} \end{tabular}} &
{\begin{tabular}[c]{@{}c@{}}\textbf{\#MOCK\_} \\ \textbf{OBJECTS} \end{tabular}} &
{\begin{tabular}[c]{@{}c@{}}\textbf{\#MOCK\_} \\ \textbf{METHODS} \end{tabular}} &
\textbf{\#STUBS} & 
{\begin{tabular}[c]{@{}c@{}}\textbf{\#OO\_} \\ \textbf{STMNTS} \end{tabular}} &
{\begin{tabular}[c]{@{}c@{}}\textbf{\#PO\_} \\ \textbf{STMNTS} \end{tabular}} &
{\begin{tabular}[c]{@{}c@{}}\textbf{\#CO\_} \\ \textbf{STMNTS} \end{tabular}} &
{\begin{tabular}[c]{@{}c@{}}\textbf{\#SUCCESSFULLY\_} \\ \textbf{MIMIC} \end{tabular}} &
{\begin{tabular}[c]{@{}c@{}}\textbf{\#INCOMPLETELY\_} \\ \textbf{MIMIC} \end{tabular}} &
{\begin{tabular}[c]{@{}c@{}}\textbf{\#UNHANDLED\_} \\ \textbf{MUT\_BEHAVIOR} \end{tabular}} \\ \hline
\rowcolor[HTML]{E9EAEB} 
MUT \# 1 & 7 & 1 & 2 & 41 B &
1 & 1 & 0 & 0 & 1 & 1 & 
0 & 0 & 2 \\ 
MUT \# 2 & 19 & 1 & 3 & 47 B &
1 & 3 & 6 & 1 & 6 & 9 &
0 & 3 & 0 \\ 
\rowcolor[HTML]{E9EAEB} 
MUT \# 3 & 8 & 1 & 2 & 53 B &
1 & 2 & 0 & 0 & 2 & 3 & 
2 & 0 & 0 \\
MUT \# 4 & 5 & 1 & 2 & 284 B &
1 & 1 & 1 & 0 & 1 & 1 & 
2 & 0 & 0 \\ 
\rowcolor[HTML]{E9EAEB} 
MUT \# 5 & 15 & 1 & 3 & 477 B &
1 & 1 & 0 & 1 & 1 & 1 & 
3 & 0 & 0 \\ 
MUT \# 6 & 13 & 0 & 3 & 982 B &
1 & 3 & 2 & 1 & 3 & 3 & 
3 & 0 & 0 \\ 
\rowcolor[HTML]{E9EAEB} 
MUT \# 7 & 7 & 1 & 2 & 1.1 KB &
1 & 1 & 1 & 1 & 1 & 1 &
2 & 0 & 0 \\ 
MUT \# 8 & 15 & 2 & 3 & 1.5 KB &
1 & 2 & 4 & 1 & 4 & 2 & 
3 & 0 & 0 \\ 
\rowcolor[HTML]{E9EAEB} 
MUT \# 9 & 25 & 0 & 3 & 2.7 KB &
2 & 5 & 9 & 1 & 10 & 8 & 
0 & 3 & 0 \\
MUT \# 10 & 15 & 0 & 3 & 2.7 KB &
2 & 3 & 11 & 1 & 7 & 6 & 
0 & 3 & 0 \\ 
\rowcolor[HTML]{E9EAEB} 
MUT \# 11 & 26 & 0 & 3 & 3 KB &
2 & 2 & 11 & 1 & 11 & 2 & 
0 & 3 & 0 \\
MUT \# 12 & 5 & 1 & 3 & 3.1 KB &
1 & 1 & 1 & 1 & 1 & 1 &
3 & 0 & 0 \\ 
\rowcolor[HTML]{E9EAEB} 
MUT \# 13 & 5 & 1 & 3 & 3.1 KB &
1 & 1 & 1 & 1 & 1 & 1 &
3 & 0 & 0 \\ 
MUT \# 14 & 25 & 0 & 3 & 56.7 KB &
2 & 4 & 5 & 1 & 5 & 5 & 
0 & 3 & 0 \\ 
\rowcolor[HTML]{E9EAEB} 
MUT \# 15 & 5 & 1 & 3 & 156.8 KB &
1 & 1 & 1 & 1 & 1 & 1 & 
3 & 0 & 0 \\ 
MUT \# 16 & 58 & 1 & 3 & 156.8 KB &
1 & 1 & 1 & 1 & 1 & 1 &
3 & 0 & 0 \\ 
\rowcolor[HTML]{E9EAEB} 
MUT \# 17 & 18 & 1 & 3 & 156.9 KB &
1 & 1 & 1 & 1 & 1 & 1 &
3 & 0 & 0 \\ 
MUT \# 18 & 5 & 1 & 3 & 383 KB &
1 & 1 & 1 & 1 & 1 & 1 &
3 & 0 & 0 \\ 
\rowcolor[HTML]{E9EAEB} 
MUT \# 19 & 74 & 1 & 2 & 590 KB &
2 & 2 & 0 & 0 & 2 & 2 & 
2 & 0 & 0 \\ 
MUT \# 20 & 8 & 0 & 2 & 729.2 KB &
1 & 2 & 20 & 0 & 20 & 20 & 
0 & 2 & 0 \\ 
\rowcolor[HTML]{E9EAEB} 
MUT \# 21 & 57 & 0 & 3 & 2.2 MB &
3 & 3 & 2 & 1 & 3 & 3 & 
0 & 0 & 3 \\ 
MUT \# 22 & 37 & 2 & 3 & 9.8 MB &
1 & 3 & 3 & 1 & 4 & 4 & 
2 & 1 & 0 \\ 
\rowcolor[HTML]{E9EAEB} 
MUT \# 23 & 4 & 0 & 2 & 9.8 MB &
2 & 1 & 0 & 0 & 1 & 1 & 
0 & 2 & 0 \\ 
\hline
\rowcolor{blue!5}
\textbf{TOTAL: 23} &
{\begin{tabular}[c]{@{}c@{}}\textbf{MEDIAN:} \\ \textbf{15} \end{tabular}} &
{\begin{tabular}[c]{@{}c@{}}\textbf{MEDIAN:} \\ \textbf{1} \end{tabular}} &
\textbf{62} & 
{\begin{tabular}[c]{@{}c@{}}\textbf{MEDIAN:} \\ \textbf{3.1 KB} \end{tabular}} &
\textbf{31} &
\textbf{45} &
\textbf{81} &
\textbf{16} & 
\textbf{88} & 
\textbf{78} &
\textbf{37} &
\textbf{20} &
\textbf{5} \\ \hline
\end{tabular}}
\end{table*}


\begin{table*}
\renewcommand*{\arraystretch}{1.6}
\centering
\caption{Experimental results for \gephi}\label{tab:gephi-results}
\resizebox{\textwidth}{!}{
\begin{tabular}{r|l|l|l?l|l|l|l|l|l|l?l|l|l}
\hline
\rowcolor{green!5}
\multicolumn{4}{c?}{\textbf{\textsc{RQ1: Method Under Test}}} &
\multicolumn{7}{c?}{\textbf{\textsc{RQ2: Production-based Mocks}}} &
\multicolumn{3}{c}{\textbf{\textsc{RQ3: Mimicking Production}}} \\ \hline
\textbf{MUT\_ID} & 
\textbf{\#LOC} & 
\textbf{\#PARAMS} & 
\textbf{\#TESTS} &
{\begin{tabular}[c]{@{}c@{}}\textbf{CAPTURED\_} \\ \textbf{OBJ\_SIZE} \end{tabular}} &
{\begin{tabular}[c]{@{}c@{}}\textbf{\#MOCK\_} \\ \textbf{OBJECTS} \end{tabular}} &
{\begin{tabular}[c]{@{}c@{}}\textbf{\#MOCK\_} \\ \textbf{METHODS} \end{tabular}} &
\textbf{\#STUBS} & 
{\begin{tabular}[c]{@{}c@{}}\textbf{\#OO\_} \\ \textbf{STMNTS} \end{tabular}} &
{\begin{tabular}[c]{@{}c@{}}\textbf{\#PO\_} \\ \textbf{STMNTS} \end{tabular}} &
{\begin{tabular}[c]{@{}c@{}}\textbf{\#CO\_} \\ \textbf{STMNTS} \end{tabular}} &
{\begin{tabular}[c]{@{}c@{}}\textbf{\#SUCCESSFULLY\_} \\ \textbf{MIMIC} \end{tabular}} &
{\begin{tabular}[c]{@{}c@{}}\textbf{\#INCOMPLETELY\_} \\ \textbf{MIMIC} \end{tabular}} &
{\begin{tabular}[c]{@{}c@{}}\textbf{\#UNHANDLED\_} \\ \textbf{MUT\_BEHAVIOR} \end{tabular}} \\ \hline
\rowcolor[HTML]{E9EAEB} 
MUT \# 1 & 16 & 1 & 2 & 48 B &
1 & 2 & 2 & 0 & 2 & 2 &
0 & 0 & 2 \\ 
MUT \# 2 & 14 & 1 & 2 & 48 B & 
1 & 1 & 1 & 0 & 1 & 1 &
0 & 0 & 2 \\ 
\rowcolor[HTML]{E9EAEB} 
MUT \# 3 & 9 & 3 & 2 & 56 B & 
1 & 1 & 1 & 0 & 1 & 1 &
2 & 0 & 0 \\ 
MUT \# 4 & 6 & 1 & 3 & 56 B & 
1 & 1 & 1 & 1 & 1 & 1 & 
0 & 0 & 3 \\ 
\rowcolor[HTML]{E9EAEB} 
MUT \# 5 & 16 & 1 & 3 & 56 B & 
1 & 1 & 1 & 1 & 1 & 1 &
3 & 0 & 0 \\ 
MUT \# 6 & 31 & 3 & 3 & 56 B 
& 1 & 1 & 0 & 1 & 1 & 1 &
3 & 0 & 0 \\ 
\rowcolor[HTML]{E9EAEB} 
MUT \# 7 & 12 & 2 & 2 & 118 B 
& 1 & 4 & 2 & 0 & 4 & 4 &
2 & 0 & 0 \\ 
MUT \# 8 & 12 & 2 & 2 & 134 B &
1 & 2 & 0 & 0 & 2 & 2 & 
1 & 1 & 0 \\ 
\rowcolor[HTML]{E9EAEB} 
MUT \# 9 & 11 & 2 & 3 & 143 B & 
1 & 2 & 2 & 1 & 2 & 2 &
3 & 0 & 0 \\ 
MUT \# 10 & 8 & 1 & 2 & 186 B 
& 1 & 1 & 0 & 0 & 1 & 1 &
0 & 0 & 2 \\
\rowcolor[HTML]{E9EAEB} 
MUT \# 11 & 18 & 3 & 2 & 206 B &
1 & 4 & 4 & 0 & 4 & 20 & 
1 & 1 & 0 \\
MUT \# 12 & 13 & 1 & 2 & 237 B &
1 & 2 & 1 & 0 & 2 & 2 & 
0 & 0 & 2 \\ 
\rowcolor[HTML]{E9EAEB} 
MUT \# 13 & 14 & 1 & 2 & 238 B &
1 & 1 & 1 & 0 & 1 & 1 & 
0 & 0 & 2 \\ 
MUT \# 14 & 12 & 1 & 2 & 255 B &
1 & 1 & 1 & 0 & 1 & 1 & 
0 & 0 & 2 \\ 
\rowcolor[HTML]{E9EAEB} 
MUT \# 15 & 17 & 1 & 2 & 291 B &
1 & 1 & 1 & 0 & 1 & 1 & 
0 & 0 & 2 \\
MUT \# 16 & 24 & 1 & 2 & 299 B &
1 & 1 & 1 & 0 & 1 & 1 &
0 & 0 & 2 \\ 
\rowcolor[HTML]{E9EAEB} 
MUT \# 17 & 43 & 3 & 3 & 307 B &
1 & 1 & 1 & 1 & 1 & 1 & 
0 & 0 & 3 \\ 
MUT \# 18 & 18 & 1 & 2 & 340 B &
1 & 1 & 1 & 0 & 1 & 1 & 
0 & 0 & 2 \\ 
\rowcolor[HTML]{E9EAEB} 
MUT \# 19 & 7 & 1 & 2 & 344 B &
1 & 1 & 1 & 0 & 1 & 1 & 
2 & 0 & 0 \\ 
MUT \# 20 & 29 & 1 & 2 & 425 B &
1 & 1 & 1 & 0 & 1 & 1 & 
0 & 0 & 2 \\ 
\rowcolor[HTML]{E9EAEB} 
MUT \# 21 & 11 & 1 & 3 & 472 B &
1 & 2 & 2 & 1 & 2 & 2 &
3 & 0 & 0 \\ 
MUT \# 22 & 40 & 3 & 2 & 1.7 KB &
3 & 3 & 1 & 0 & 3 & 3 & 
0 & 2 & 0 \\ 
\rowcolor[HTML]{E9EAEB} 
MUT \# 23 & 50 & 2 & 2 & 2.2 KB &
1 & 2 & 1 & 0 & 2 & 2 & 
0 & 2 & 0 \\
MUT \# 24 & 23 & 3 & 2 & 4.4 KB &
1 & 1 & 1 & 0 & 1 & 1 & 
2 & 0 & 0 \\ 
\rowcolor[HTML]{E9EAEB} 
MUT \# 25 & 13 & 1 & 2 & 58.2 KB &
3 & 3 & 3 & 0 & 3 & 3 & 
0 & 2 & 0 \\ 
MUT \# 26 & 4 & 1 & 3 & 58.2 KB & 
1 & 1 & 1 & 1 & 1 & 1 & 
3 & 0 & 0 \\ 
\rowcolor[HTML]{E9EAEB} 
MUT \# 27 & 23 & 3 & 2 & 59.2 KB &
1 & 1 & 1 & 0 & 1 & 1 & 
2 & 0 & 0 \\ 
MUT \# 28 & 53 & 1 & 3 & 60.1 KB &
1 & 4 & 10 & 1 & 4 & 27 & 
0 & 3 & 0 \\ 
\rowcolor[HTML]{E9EAEB} 
MUT \# 29 & 21 & 2 & 2 & 79 KB &
1 & 1 & 1 & 0 & 1 & 1 & 
2 & 0 & 0 \\ 
MUT \# 30 & 85 & 4 & 2 & 100 KB & 
1 & 1 & 1 & 0 & 1 & 1 & 
0 & 0 & 2 \\ 
\rowcolor[HTML]{E9EAEB} 
MUT \# 31 & 85 & 1 & 2 & 112.5 KB &
4 & 4 & 5 & 0 & 4 & 6 & 
0 & 0 & 2 \\ 
MUT \# 32 & 25 & 1 & 2 & 112.5 KB &
1 & 1 & 1 & 0 & 1 & 1 & 
2 & 0 & 0 \\ 
\rowcolor[HTML]{E9EAEB} 
MUT \# 33 & 10 & 1 & 3 & 149.5 KB &
1 & 1 & 1 & 1 & 1 & 1 &
2 & 1 & 0 \\ 
MUT \# 34 & 90 & 0 & 3 & 177.4 KB &
1 & 1 & 1 & 1 & 1 & 1 &
0 & 3 & 0 \\ 
\rowcolor[HTML]{E9EAEB} 
MUT \# 35 & 15 & 3 & 2 & 196.3 KB &
1 & 3 & 0 & 0 & 3 & 3 & 
2 & 0 & 0 \\ 
MUT \# 36 & 28 & 2 & 2 & 276 KB &
1 & 1 & 1 & 0 & 1 & 1 & 
0 & 0 & 2 \\ 
\rowcolor[HTML]{E9EAEB} 
MUT \# 37 & 40 & 2 & 2 & 300 KB &
1 & 1 & 1 & 0 & 1 & 1 & 
0 & 0 & 2 \\ 
MUT \# 38 & 18 & 6 & 2 & 370 KB &
1 & 1 & 1 & 0 & 1 & 1 & 
0 & 0 & 2 \\ 
\rowcolor[HTML]{E9EAEB} 
MUT \# 39 & 16 & 6 & 3 & 600 KB &
1 & 1 & 1 & 1 & 1 & 1 &
0 & 0 & 3 \\ 
MUT \# 40 & 5 & 2 & 2 & 3.6 MB &
1 & 1 & 1 & 0 & 1 & 1 & 
0 & 0 & 2 \\ 
\rowcolor[HTML]{E9EAEB} 
MUT \# 41 & 62 & 6 & 2 & 3.6 MB &
2 & 8 & 8 & 0 & 8 & 8 & 
0 & 0 & 2 \\ 
MUT \# 42 & 15 & 6 & 2 & 3.8 MB &
1 & 1 & 1 & 0 & 1 & 1 & 
0 & 0 & 2 \\ 
\rowcolor[HTML]{E9EAEB} 
MUT \# 43 & 26 & 0 & 2 & 3.9 MB &
1 & 1 & 6 & 0 & 1 & 1 & 
0 & 2 & 0 \\ 
MUT \# 44 & 185 & 0 & 2 & 3.9 MB &
1 & 1 & 6 & 0 & 1 & 1 & 
0 & 0 & 2 \\ 
\rowcolor[HTML]{E9EAEB} 
MUT \# 45 & 11 & 0 & 2 & 3.9 MB &
1 & 1 & 0 & 0 & 1 & 1 & 
2 & 0 & 0 \\ 
MUT \# 46 & 123 & 0 & 2 & 3.9 MB &
2 & 3 & 1 & 0 & 3 & 3 & 
0 & 2 & 0 \\ 
\rowcolor[HTML]{E9EAEB} 
MUT \# 47 & 59 & 5 & 2 & 4 MB &
1 & 1 & 3 & 0 & 1 & 1 & 
0 & 0 & 2 \\ 
MUT \# 48 & 45 & 6 & 2 & 4.1 MB &
2 & 2 & 3 & 0 & 2 & 2 & 
0 & 0 & 2 \\ 
\rowcolor[HTML]{E9EAEB} 
MUT \# 49 & 20 & 0 & 2 & 4.2 MB &
1 & 1 & 0 & 0 & 1 & 1 & 
2 & 0 & 0 \\ 
MUT \# 50 & 18 & 2 & 3 & 4.5 MB &
1 & 1 & 0 & 1 & 1 & 1 & 
0 & 0 & 3 \\
\rowcolor[HTML]{E9EAEB} 
MUT \# 51 & 126 & 0 & 2 & 10.8 MB &
1 & 1 & 6 & 0 & 1 & 1 & 
0 & 2 & 0 \\ 
MUT \# 52 & 24 & 0 & 2 & 10.8 MB &
1 & 1 & 1 & 0 & 1 & 1 & 
0 & 2 & 0 \\ 
\rowcolor[HTML]{E9EAEB} 
MUT \# 53 & 24 & 0 & 2 & 10.8 MB &
1 & 1 & 1 & 0 & 1 & 1 & 
0 & 2 & 0 \\ 
MUT \# 54 & 15 & 2 & 2 & 11.2 MB &
1 & 1 & 1 & 0 & 1 & 1 & 
2 & 0 & 0 \\ 
\rowcolor[HTML]{E9EAEB} 
MUT \# 55 & 11 & 2 & 2 & 11.6 MB &
1 & 2 & 4 & 0 & 2 & 2 & 
1 & 1 & 0 \\ 
MUT \# 56 & 7 & 3 & 2 & 11.6 MB &
1 & 1 & 1 & 0 & 1 & 1 & 
2 & 0 & 0 \\ 
\rowcolor[HTML]{E9EAEB} 
MUT \# 57 & 32 & 3 & 2 & 11.6 MB &
1 & 2 & 3 & 0 & 3 & 3 & 
0 & 0 & 2 \\ 
\hline
\rowcolor{green!5}
\textbf{TOTAL: 57} &
{\begin{tabular}[c]{@{}c@{}}\textbf{MEDIAN:} \\ \textbf{18} \end{tabular}} &
{\begin{tabular}[c]{@{}c@{}}\textbf{MEDIAN:} \\ \textbf{1} \end{tabular}} &
\textbf{126} & 
{\begin{tabular}[c]{@{}c@{}}\textbf{MEDIAN:} \\ \textbf{60.1 KB} \end{tabular}} &
\textbf{67} &
\textbf{93} &
\textbf{103} &
\textbf{12} & 
\textbf{94} & 
\textbf{135} &
\textbf{44} &
\textbf{26} &
\textbf{56} \\ \hline
\end{tabular}}
\end{table*}

\begin{table*}
\renewcommand*{\arraystretch}{1.6}
\centering
\caption{Experimental results for \pdfbox}\label{tab:pdfbox-results}
\resizebox{\textwidth}{!}{
\begin{tabular}{r|l|l|l?l|l|l|l|l|l|l?l|l|l}
\hline
\rowcolor{yellow!5}
\multicolumn{4}{c?}{\textbf{\textsc{RQ1: Method Under Test}}} &
\multicolumn{7}{c?}{\textbf{\textsc{RQ2: Production-based Mocks}}} &
\multicolumn{3}{c}{\textbf{\textsc{RQ3: Mimicking Production}}} \\
\hline
\textbf{MUT\_ID} & 
\textbf{\#LOC} & 
\textbf{\#PARAMS} & 
\textbf{\#TESTS} & 
{\begin{tabular}[c]{@{}c@{}}\textbf{CAPTURED\_} \\ \textbf{OBJ\_SIZE} \end{tabular}} &
{\begin{tabular}[c]{@{}c@{}}\textbf{MOCK\_} \\ \textbf{OBJECTS} \end{tabular}} &
{\begin{tabular}[c]{@{}c@{}}\textbf{MOCK\_} \\ \textbf{METHODS} \end{tabular}} &
\textbf{\#STUBS} & 
{\begin{tabular}[c]{@{}c@{}}\textbf{\#OO\_} \\ \textbf{STMNTS} \end{tabular}} &
{\begin{tabular}[c]{@{}c@{}}\textbf{\#PO\_} \\ \textbf{STMNTS} \end{tabular}} &
{\begin{tabular}[c]{@{}c@{}}\textbf{\#CO\_} \\ \textbf{STMNTS} \end{tabular}} &
{\begin{tabular}[c]{@{}c@{}}\textbf{\#SUCCESSFULLY\_} \\ \textbf{MIMIC} \end{tabular}} &
{\begin{tabular}[c]{@{}c@{}}\textbf{\#INCOMPLETELY\_} \\ \textbf{MIMIC} \end{tabular}} &
{\begin{tabular}[c]{@{}c@{}}\textbf{\#UNHANDLED\_} \\ \textbf{MUT\_BEHAVIOR} \end{tabular}} \\ \hline
\rowcolor[HTML]{E9EAEB} 
MUT \# 1 & 84 & 5 & 2 & 37 B &
1 & 1 & 0 & 0 & 1 & 1 & 
2 & 0 & 0 \\ 
MUT \# 2 & 6 & 1 & 2 & 352 B &
1 & 2 & 2 & 0 & 2 & 4 & 
0 & 0 & 2 \\  
\rowcolor[HTML]{E9EAEB} 
MUT \# 3 & 11 & 1 & 2 & 3.3 KB &
1 & 1 & 0 & 0 & 1 & 1 & 
0 & 0 & 2 \\  
MUT \# 4 & 5 & 1 & 2 & 4.3 KB &
1 & 1 & 0 & 0 & 1 & 1 & 
2 & 0 & 0 \\  
\rowcolor[HTML]{E9EAEB} 
MUT \# 5 & 5 & 1 & 2 & 4.7 KB &
1 & 1 & 0 & 0 & 1 & 1 & 
2 & 0 & 0 \\  
MUT \# 6 & 101 & 3 & 2 & 6.5 KB &
1 & 1 & 1 & 0 & 1 & 1 & 
0 & 0 & 2 \\ 
\rowcolor[HTML]{E9EAEB} 
MUT \# 7 & 28 & 1 & 2 & 11.7 KB &
1 & 1 & 0 & 0 & 5 & 1 &
2 & 0 & 0 \\  
MUT \# 8 & 34 & 1 & 2 & 267 KB &
1 & 1 & 1 & 0 & 3 & 2 &
2 & 0 & 0 \\  
\rowcolor[HTML]{E9EAEB} 
MUT \# 9 & 42 & 2 & 3 & 1.6 MB &
1 & 1 & 1 & 1 & 1 & 1 &
3 & 0 & 0 \\  
MUT \# 10 & 33 & 1 & 2 & 2 MB &
1 & 1 & 0 & 0 & 1 & 1 & 
2 & 0 & 0 \\  
\rowcolor[HTML]{E9EAEB} 
MUT \# 11 & 16 & 1 & 2 & 2.2 MB &
1 & 1 & 1 & 0 & 1 & 1 & 
0 & 0 & 2 \\  
MUT \# 12 & 10 & 0 & 3 & 2.3 MB &
1 & 2 & 2 & 1 & 2 & 2 &
3 & 0 & 0 \\  
\rowcolor[HTML]{E9EAEB} 
MUT \# 13 & 19 & 3 & 3 & 2.3 MB &
1 & 2 & 2 & 1 & 2 & 2 &
3 & 0 & 0 \\  
MUT \# 14 & 19 & 0 & 2 & 2.4 MB &
1 & 1 & 1 & 0 & 1 & 1 & 
2 & 0 & 0 \\  
\rowcolor[HTML]{E9EAEB} 
MUT \# 15 & 18 & 0 & 3 & 3 MB &
1 & 1 & 1 & 1 & 1 & 1 &
3 & 0 & 0 \\ 
MUT \# 16 & 18 & 2 & 2 & 3 MB &
1 & 1 & 1 & 0 & 1 & 1 & 
2 & 0 & 0 \\  
\rowcolor[HTML]{E9EAEB} 
MUT \# 17 & 8 & 0 & 2 & 3.2 MB &
1 & 1 & 1 & 0 & 1 & 1 & 
2 & 0 & 0 \\  
MUT \# 18 & 9 & 1 & 3 & 3.6 MB &
1 & 1 & 1 & 1 & 1 & 1 &
3 & 0 & 0 \\  
\rowcolor[HTML]{E9EAEB}
MUT \# 19 & 8 & 0 & 3 & 3.8 MB &
1 & 1 & 1 & 1 & 1 & 1 &
3 & 0 & 0 \\  
MUT \# 20 & 18 & 1 & 2 & 3.8 MB &
1 & 2 & 2 & 0 & 2 & 2 & 
2 & 0 & 0 \\  
\rowcolor[HTML]{E9EAEB}
MUT \# 21 & 5 & 1 & 2 & 4.4 MB &
1 & 1 & 1 & 0 & 1 & 1 & 
2 & 0 & 0 \\  
MUT \# 22 & 19 & 0 & 3 & 4.6 MB &
1 & 2 & 2 & 1 & 2 & 2 &
3 & 0 & 0 \\  
\rowcolor[HTML]{E9EAEB}
MUT \# 23 & 39 & 1 & 2 & 4.6 MB &
1 & 1 & 0 & 0 & 1 & 1 & 
2 & 0 & 0 \\  
MUT \# 24 & 85 & 1 & 3 & 5.3 MB &
3 & 1 & 1 & 1 & 1 & 1 & 
0 & 3 & 0 \\  
\rowcolor[HTML]{E9EAEB}
MUT \# 25 & 40 & 1 & 2 & 5.3 MB &
1 & 1 & 1 & 0 & 1 & 1 & 
0 & 0 & 2 \\  
MUT \# 26 & 328 & 2 & 2 & 5.4 MB &
1 & 1 & 1 & 0 & 1 & 1 & 
0 & 0 & 2 \\  
\rowcolor[HTML]{E9EAEB}
MUT \# 27 & 11 & 1 & 3 & 5.6 MB &
1 & 2 & 2 & 1 & 2 & 2 &
3 & 0 & 0 \\ 
MUT \# 28 & 29 & 2 & 3 & 5.8 MB &
1 & 2 & 3 & 1 & 2 & 4 & 
2 & 1 & 0 \\  
\rowcolor[HTML]{E9EAEB}
MUT \# 29 & 79 & 4 & 2 & 6.9 MB &
1 & 1 & 0 & 0 & 1 & 1 & 
2 & 0 & 0 \\ 
MUT \# 30 & 19 & 1 & 2 & 6.4 MB &
1 & 1 & 0 & 0 & 1 & 1 & 
2 & 0 & 0 \\  
\rowcolor[HTML]{E9EAEB}
MUT \# 31 & 31 & 1 & 2 & 6.6 MB &
1 & 1 & 0 & 0 & 1 & 1 & 
2 & 0 & 0 \\  
MUT \# 32 & 62 & 4 & 2 & 6.8 MB &
1 & 1 & 1 & 0 & 1 & 1 & 
2 & 0 & 0 \\  
\rowcolor[HTML]{E9EAEB} 
MUT \# 33 & 23 & 1 & 2 & 7.2 MB &
1 & 1 & 0 & 0 & 1 & 1 & 
2 & 0 & 0 \\ 
MUT \# 34 & 27 & 2 & 2 & 7.4 MB &
1 & 1 & 0 & 0 & 1 & 1 & 
0 & 0 & 2 \\  
\rowcolor[HTML]{E9EAEB}
MUT \# 35 & 40 & 1 & 2 & 7.6 MB &
1 & 1 & 1 & 0 & 1 & 1 & 
0 & 0 & 2 \\  
MUT \# 36 & 20 & 1 & 2 & 8.3 MB &
1 & 1 & 1 & 0 & 1 & 1 & 
0 & 2 & 0 \\  
\rowcolor[HTML]{E9EAEB}
MUT \# 37 & 12 & 1 & 2 & 9.5 MB &
1 & 1 & 0 & 0 & 1 & 1 & 
2 & 0 & 0 \\ 
MUT \# 38 & 12 & 1 & 2 & 9.5 MB &
1 & 1 & 0 & 0 & 1 & 1 & 
2 & 0 & 0 \\ 
\rowcolor[HTML]{E9EAEB} 
MUT \# 39 & 11 & 1 & 2 & 9.7 MB & 
1 & 1 & 0 & 0 & 1 & 1 & 
2 & 0 & 0 \\  
MUT \# 40 & 23 & 1 & 2 & 10.1 MB &
1 & 1 & 0 & 0 & 1 & 0 & 
2 & 0 & 0 \\ 
\rowcolor[HTML]{E9EAEB}
MUT \# 41 & 11 & 0 & 2 & 11.1 MB &
1 & 1 & 0 & 0 & 1 & 1 & 
1 & 1 & 0 \\ 
MUT \# 42 & 5 & 1 & 2 & 21.4 MB &
1 & 1 & 0 & 0 & 1 & 1 & 
2 & 0 & 0 \\  
\rowcolor[HTML]{E9EAEB} 
MUT \# 43 & 8 & 4 & 2 & 30.8 MB &
3 & 5 & 4 & 0 & 7 & 13 & 
0 & 2 & 0 \\ 
MUT \# 44 & 14 & 0 & 2 & 31 MB &
1 & 2 & 0 & 0 & 2 & 2 & 
0 & 0 & 2 \\ 
\rowcolor[HTML]{E9EAEB} 
MUT \# 45 & 17 & 2 & 2 & 32 MB &
1 & 2 & 0 & 0 & 3 & 3 & 
0 & 2 & 0 \\ 
MUT \# 46 & 36 & 2 & 2 & 32 MB &
1 & 1 & 0 & 0 & 1 & 1 & 
0 & 0 & 2 \\ 
\rowcolor[HTML]{E9EAEB}
MUT \# 47 & 12 & 0 & 2 & 34 MB &
2 & 5 & 0 & 0 & 5 & 5 & 
2 & 0 & 0 \\  
MUT \# 48 & 117 & 1 & 2 & 39 MB &
1 & 2 & 2 & 0 & 2 & 3 & 
0 & 0 & 2 \\  
\hline
\rowcolor{yellow!5}
\textbf{TOTAL: 48} &
{\begin{tabular}[c]{@{}c@{}}\textbf{MEDIAN:} \\ \textbf{19} \end{tabular}} &
{\begin{tabular}[c]{@{}c@{}}\textbf{MEDIAN:} \\ \textbf{1} \end{tabular}} &
\textbf{106} & 
{\begin{tabular}[c]{@{}c@{}}\textbf{MEDIAN:} \\ \textbf{5.3 MB} \end{tabular}} &
\textbf{53} &
\textbf{66} &
\textbf{38} &
\textbf{10} & 
\textbf{75} & 
\textbf{80} &
\textbf{73} &
\textbf{11} &
\textbf{22} \\ \hline
\end{tabular}}
\end{table*}


\subsection{Results for RQ1 [Methods Under Test]}\label{sec:results-rq1}
As presented in the first four columns of \autoref{tab:graphhopper-results}, \autoref{tab:gephi-results}, and \autoref{tab:pdfbox-results}, \rick generates tests for $23$ MUTs in \graphhopper, $57$ in \gephi, and $48$ in \pdfbox. 
In total, \rick generates tests for $128$ MUTs which have at least one mockable method call. The median number of LOCs in these $128$ target methods is $18$, while the largest method is MUT\#26 in \pdfbox with $328$ lines of code. The median number of parameters for the MUTs is $1$, while several MUTs (such as MUT\#38 and MUT\#39 in \gephi) take as many as $6$ parameters. In general, \rick handles a wide variety of MUTs in the case studies, with successful identification and instrumentation of these methods, detailed monitoring in production, as well as the generation of tests that compile and run. 

These results validate that mock generation from production can indeed be fully automated, and is robust with respect to the complexity of real world methods. \rick handles the diversity of methods, data types, and interactions observed in real software and production usage scenarios.

In \autoref{tab:production-usage}, we see that the workloads trigger the execution of  $72$ MUTs in \graphhopper, $68$ in \gephi, and $72$ in \pdfbox. A subset of them are actually used as targets for test generation: $23$, $57$, and $48$ MUTs in the respective case studies. This happens due to two reasons. 

First, when statically identifying targets for test generation, \rick finds MUTs with mockable method calls. However, as highlighted in \autoref{sec:rick-monitoring}, MUTs may be invoked, without their corresponding mockable methods being called, because of the flow of control through the program.
For example, a mockable method call may happen only within a certain path through the MUT, which is not observed in production. In this case, a test with mocks is not generated. Second, the receiving objects for some MUTs are sometimes too large and complex to be captured through serialization, which is the case for some MUTs invoked within \graphhopper. 
We have observed serialized object snapshots beyond tens of megabytes, which reaches the scalability limits of state of the art serialization techniques.

The difference between the number of invoked MUTs and the number of MUTs for which \rick generates tests is most significant for \graphhopper. We notice that many receiving objects for \graphhopper do not get successfully serialized owing to their large size. For example, the receiving object for an invoked MUT was as large as $454$ MB, before even being serialized as XML. To counter this, we allocated more heap space, increasing it up to 9 GB, while deploying the server, yet were unsuccessful in serializing it. 
Our research on using serialization for automated mocking clearly touches the frontier of serialization for handling arbitrarily large and complex data from production.

The total number of tests generated by \rick for the $128$ target MUTs is $294$. Recall from \autoref{sec:production-usage} that we only monitor and generate tests for a single invocation -- the first one -- of each of the target MUTs. Note also that, as signified by the column \#TESTS in \autoref{tab:graphhopper-results}, \autoref{tab:gephi-results}, and \autoref{tab:pdfbox-results}, \rick generates either $2$ or $3$ tests for each target MUT. The number of tests generated for an MUT depends on its return type. For MUTs that return a non-void, primitive value, such as MUT\#2 in \graphhopper, \rick generates an \OO test to assert on the output of the MUT invocation, in addition to a \PO and \CO test. Across the three case studies, \rick generates \OO tests for $38$ MUTs, and \PO and \CO tests for all $128$ MUTs.

\begin{mdframed}[style=mpdframe,nobreak=true,frametitle=Answer to RQ1]
\rick captures the production behavior for a set of $128$ out of the $212$ MUTs invoked in production. \rick transforms the data collected from these production invocations into $294$ concrete tests with different oracles. The key result of RQ1 is that \rick handles a large variety of real world methods in an end to end manner, from monitoring in production to the generation of tests with mocks.
\end{mdframed}

\subsection{Results for RQ2 [Production-based Mocks]}\label{sec:results-rq2}

\revisedthree{\rick generates mocks from real data observed in production.
While RQ1 has demonstrated feasibility, RQ2 explores how production data is reflected in the generated test cases.
}

Columns $5$ through $11$ in \autoref{tab:graphhopper-results}, \autoref{tab:gephi-results}, and \autoref{tab:pdfbox-results} present the results for RQ2. Each row highlights the data that \rick collects for one MUT and its mock method call(s) from production.
The column CAPTURED\_OBJ\_SIZE presents the size on disk (in B / KB / MB) of the serialized object states for the receiving object and, if present, the parameter objects of the MUT. We also present the number of external objects mocked within the test (\#MOCK\_OBJECTS), the number of methods called on these mock objects (\#MOCK\_METHODS), the number of stubs (\#STUBS) defined for these mock methods, and the number of statements corresponding to the \OO, \PO, and \CO oracles in the generated tests (signified through \#OO\_STMNTS, \#PO\_STMNTS, and \#CO\_STMNTS, respectively).

\revisedthree{We first discuss the serialized production state, defined as a receiving object on which the MUT is invoked, as well as the parameters with which this invocation is made.
Recall that the receiving object is serialized. 
For example, consider MUT\#32 in \pdfbox (\autoref{tab:pdfbox-results}).
\rick captures the production state of the receiving object on which MUT\#32 gets invoked, and the $4$ parameters for this invocation.
The size of these captured objects amounts to a total of $6.8$ MB.
These objects serve as a snapshot for reproducing the production state within the $2$ tests generated by \rick for MUT\#32.
For our experiments with the three applications, the maximum size of the captured objects is $39$ MB for MUT\#48 in \pdfbox.
The median size of these captured objects is $486$ bytes of serialized data.
\autoref{tab:pdfbox-results} characterizes the realistic nature of the objects that \rick captures over  $128$ real-world MUTs, as they are invoked during the end-to-end execution of the applications.}

The main feature of \rick is to monitor and collect data about mockable method calls that occur within the MUT, with the receiving objects and passed parameters. Specifically, as detailed in \autoref{sec:rick-design}, \rick mocks the parameters and fields of external types, on which mockable method calls occur in production.
For example, there are $2$ mock objects within the $2$ tests generated for MUT\#19 in \graphhopper (\autoref{tab:graphhopper-results}) for two fields in the declaring type of MUT\#19. Moreover, $2$ mock method calls are made within the tests, one on each of the $2$ mock objects. 
In comparison, there is $1$ mock object within each of the $3$ tests generated for MUT\#21 in \gephi (\autoref{tab:gephi-results}). This mock object replaces the parameter of MUT\#21, on which $2$ mockable method calls are observed by \rick in production. 
In total, \rick uses $151$ mock objects as parameter or field across the generated tests. $204$ mock methods are invoked on these mock objects, reflecting the production interactions with external objects within the MUTs.
These generated mock objects recreate actual interactions of the MUT with its environment. The data serialized from production provide developers with realistic test data.

\begin{lstlisting}[language=Java, belowskip={-10pt}, label={lst:gephi-test-po}, caption={The \PO test generated by \rick for MUT\#7 in \gephi, which is a method called \texttt{moveNode}. The test mocks a parameter on which four mocked methods are invoked. The behavior of two mocked methods is stubbed within the test.}, float]
@Test
@DisplayName("moveNode with parameter oracle, 
mocking Node.x(), Node.y(), Node.setX(float), Node.setY(float)")
public void testMoveNode_PO() throws Exception {
  // Arrange
  StepDisplacement receivingObject = deserialize("receiving.xml");
  Object[] paramObjects = deserialize("params.xml");
  ForceVector paramObject2 = (ForceVector) paramObjects[1];
  Node mockNode = Mockito.mock(Node.class);
  when(mockNode.x()).thenReturn(-423.78378F);
  when(mockNode.y()).thenReturn(107.523186F);
  
  // Act
  receivingObject.moveNode(mockNode, paramObject2);

  // Assert
  verify(mockNode, atLeastOnce()).x();
  verify(mockNode, atLeastOnce()).y();
  verify(mockNode, atLeastOnce()).setX(-403.92587F);
  verify(mockNode, atLeastOnce()).setY(105.14341F);
}
\end{lstlisting}

Within the generated tests, the behavior of the mock objects is defined through method stubs. A stub provides the canned response that should be returned from a non-void method called on a mock object, given a set of parameters, i.e., it defines the behavior of a mock method within the test for an MUT. \rick sources the parameters and the primitive returned value from production observations, and represents them through stubs within the generated tests.
For example, \rick generates $2$ tests for MUT\#7 in \gephi (\autoref{tab:gephi-results}). This MUT is the method \texttt{moveNode(Node,ForceVector)}. Within each of the generated tests, $1$ of the parameters of \texttt{moveNode} is mocked, and $4$ methods are invoked on this mock object. We present the \PO test generated for \texttt{moveNode} in \autoref{lst:gephi-test-po}. On line $9$ the \texttt{Node} object is mocked. Within the invocation of \texttt{moveNode} in production, \rick recorded the invocation of the methods \texttt{x()} and \texttt{y()} on the \texttt{Node} object, including their returned values. Consequently, \rick expresses their behavior through the $2$ stubs in the generated test (lines $10$ and $11$) using this production data. The invocations of the two other mockable methods \texttt{setX(float)} and \texttt{setY(float)} are not stubbed because they are void methods.
In total, \rick generates a total of $222$ stubs to mimic the production behavior of mockable method calls that occur within MUTs. These generated stubs guide the behavior of the MUT within the generated test, per the observations made for it in production.

We observe from the column \#OO\_STMNTS in \autoref{tab:graphhopper-results}, \autoref{tab:gephi-results}, and \autoref{tab:pdfbox-results}, that the number of \OO statements for any MUT is either $0$ or $1$. This is because the oracle in \OO tests is expressed as a single assertion statement for the output of an MUT that returns a primitive value. However, the number of \PO and \CO statements within each test varies depending on the observations made for the corresponding MUT in production. For instance, three tests are generated for MUT\#8 in \graphhopper (\autoref{tab:graphhopper-results}). In each of the three tests, there is $1$ mock object, and $2$ different mock methods are invoked on this mock object. The behavior of the $2$ mock methods is defined through the $4$ generated stubs. The \OO test has an assertion to verify the output of MUT\#8. The \PO test has $4$ verification statements to verify the parameters with which the mock methods are called. The \CO test has $2$ verification statements which correspond to the observations made by \rick about the sequence and frequency of these mock method calls within the invocation of MUT\#8 in production.

In total, across the $294$ tests generated for the $128$ MUTs, \rick generates $38$ assertion statements (one in each \OO), $257$ statements that verify the parameters with which mock methods are called, and $293$ statements to verify the sequence and frequency of mock method calls. Furthermore, \rick uses the parameters passed to, and the value returned from the mockable method calls observed in production, to generate a total of $222$ stubs across the $294$ tests.

\begin{mdframed}[style=mpdframe,nobreak=true,frametitle=Answer to RQ2]
\rick captures a wide range of production data for test generation.
We have demonstrated with static insights that it can handle well the two dimensions of mock-based testing: capturing the production state in  mock objects, and stubbing real-world methods.
The analysis of the generated tests shows that \rick can generate various types of oracles that verify different aspects of the MUT interacting with its environment. 
\end{mdframed}

\subsection{Results for RQ3 [Mimicking Production]}\label{sec:results-rq3}
The results for RQ3 are presented in the last three columns in \autoref{tab:graphhopper-results}, \autoref{tab:gephi-results}, and \autoref{tab:pdfbox-results}.
All the tests generated by \rick for the $128$ MUTs are self-contained and compile correctly.
\revisedthree{We run each generated test ten times and verify that it is not flaky.}
Next, for each test, we report the status of its execution. In each row, the column \#SUCCESSFULLY\_MIMIC highlights the number of tests that completely recreate the observed production context with mocks and oracle(s) that pass, while
\#INCOMPLETELY\_MIMIC signifies the number of tests for which at least one oracle fails. The last column, \#UNHANDLED\_MUT\_BEHAVIOR, represents those test executions where we observe a runtime exception thrown by the MUT. We now discuss the implications of each of these scenarios, and why they occur. 

An \OO test successfully mimics the production context if the MUT returns the same output as it did in production, when invoked with mocks replacing the external objects, and stubs for the behavior of the mock method calls. If the test does not completely mimic the production behavior of the MUT, the invocation of the MUT returns a different output, which is unequal to the one returned by the MUT in production. Consequently, the assertion statement within the \OO test fails.

For a \PO test to be successful, all the verification statements must pass, indicating that mock methods are called by the MUT within the generated test with the same arguments as the ones observed in production. In comparison, a failure in any of the verification statements implies that a mock method call occurs with different parameters than the ones observed for its invocation in production. This implies that the test does not faithfully recreate the observed interactions of the MUT and the mock method.

A passing \CO test verifies that the mock method calls occur in the same order and the same number of times within the MUT, as they did in production. On the contrary, if the test does not completely mimic the order and/or frequency of mock method calls, a verification statement will fail.

The proportion of tests that successfully mimic production behavior differs across case studies (column \#SUCCESSFULLY\_MIMIC). Of the tests generated for \graphhopper, $59.7\%$ are successful. In \gephi, the successful tests are $35\%$ of the total generated tests, while in \pdfbox $68.9\%$ tests are successful. Overall, the $154$ successful tests account for $52.4\%$ of all the tests generated. 
This is arguably a high ratio given the multi-stage pipeline of \rick, where each stage can fail in some conditions.
In $52.4\%$ of cases, all stages of \rick succeed: All the captured objects are correctly serialized and deserialized before the MUT is invoked, recreating an appropriate and realistic execution state. Also, the mock methods are successfully stubbed: they mimic production behavior without impacting the behavior of the MUT.

In the column \#INCOMPLETELY\_MIMIC, we observe that $57$ generated tests, which account for $19.4$\% of the total, have a failing oracle, implying that they do not completely mimic production behavior. This includes $32.2\%$ of the \graphhopper tests, $20.6\%$ \gephi tests, and $10.4\%$ of the tests generated for \pdfbox. These failures can occur due to the following reasons.

\emph{Unfaithful recreation of production state:} The captured objects may be inaccurately deserialized within the generated test, implying that production states are not completely recreated. Deserialization of complex objects captured from production is a known problem and a key challenge for recreating real production conditions in generated tests \cite{9526340,alshahwan2024observation}. A test may also fail because a production resource, such as a file, is not available during test execution, resulting in an exception. Moreover, since mocks are skeletal objects that substitute a concrete object within the test, they can induce a change in the path taken through the MUT, which renders the oracle unsuccessful. \revisedthree{For example, the tests generated for MUT\#2 in \graphhopper (\autoref{tab:graphhopper-results}) fail due to failing \OO, \PO, and \CO oracles. The tests mock an object of type \texttt{com.graphhopper.util.PointList}, and MUT\#2 tries to invoke a loop over the size of this mock list of points. Since the loop is not exercised, a different path is traversed through the MUT than the one observed in production.}

\emph{Type-based stubbing:} We observe that some tests fail because of the granularity of stubbing. 
{Glowroot, the current infrastructure for monitoring within \rick, identifies a target type based on its fully qualified name.
Consequently, \rick stubs mockable methods called on the type of an object, but not based on specific instances of the object.}
A failure can occur if an MUT calls the same mockable method on multiple parameters or fields of the same type. For instance, we present an excerpt of MUT\#24 in \pdfbox in \autoref{lst:pdfbox-mut}. This method \texttt{codeToGID(int)} (line $6$), has three calls to the same mockable method, \texttt{getGlyphId(int)} (lines $9$, $11$, and $13$). These calls are made on three different fields of type \texttt{CmapSubtable} called \texttt{cmapWinUnicode}, \texttt{cmapWinSymbol}, and \texttt{cmapMacRoman}
defined in the \texttt{PDTrueTypeFont} class (lines $2$ to $4$). \rick records the mockable method call in production, and mocks the three fields. However, the information on which of these mock fields actually calls the mock method is not available. 
One solution would be to do object-based stubbing, but we are not aware of any work on this and consider this sophistication as future work. 

\begin{lstlisting}[language=Java, belowskip={-10pt}, label={lst:pdfbox-mut}, caption={The same mockable method, \texttt{getGlyphId}, is called on three different fields of the same type within MUT\#24 of \pdfbox, \texttt{codeToGID}}, float]
class PDTrueTypeFont {
  CmapSubtable cmapWinUnicode;
  CmapSubtable cmapWinSymbol;
  CmapSubtable cmapMacRoman;
  
  public int codeToGID(int code) {
    ...
    if (...) {
      gid = cmapWinUnicode.getGlyphId(...);
    } else if (...) {
      gid = cmapMacRoman.getGlyphId(...);
    } else {
      gid = cmapWinSymbol.getGlyphId(...);
    }
    ...
    return gid;
  }
}
\end{lstlisting}

\revisedthree{We now discuss the cases of \#UNHANDLED\_MUT\_BEHAVIOR. The execution of $83$ of the $294$ generated tests ($28.2\%$) causes exceptions to be thrown by the MUT.
For example, the MUT may have multiple non-mockable interactions with the mock object, before the mock method is invoked on it.
Any of these other interactions can behave unexpectedly, resulting in exceptions to be raised before the oracle is even evaluated within the test.
Examining the test execution logs, we see such unhandled behaviors as exceptions.
We observe these cases in all three applications: $8.1\%$ in \graphhopper, $44.5\%$ in \gephi, $20.7\%$ in \pdfbox.
For example, a null pointer exception is thrown from MUT\#1 in \graphhopper, when it calls other methods on the mock object before the mockable method is called.
Across the 83 unhandled cases, 74 arise from null pointer exceptions, of which 54 are in \gephi, 18 in \pdfbox, and 2 in \graphhopper.
We find the other two unhandled cases for \gephi in the tests generated for MUT\#31, \texttt{addEdge(EdgeDraft)}, where the parameter \texttt{EdgeDraft} is one of the 4 mock objects.
This MUT calls another method, which has been designed by the developers of \gephi to throw a \texttt{ClassCastException} if \texttt{EdgeDraft} is not an instance of \texttt{ElementDraftImpl}, which fully explains the failure to execute with a mock.
In \graphhopper, the three tests generated for \texttt{load} (MUT\#21) mock the field \texttt{EncodingManager} within the type \texttt{GraphHopper}.
This MUT invokes method \texttt{checkProfilesConsistency}, which is designed to return an \texttt{IllegalArgumentException} if the \texttt{EncodingManager} does not have an encoder for the vehicle set in the profile.
In \pdfbox, the four tests generated for \texttt{prepareForDecryption} (MUT\#6) and \texttt{getColorSpace} (MUT\#34) throw an \texttt{IOException} because methods called by these MUTs find an unexpected value when accessing a field within the mock object.
Across all these cases, the unhandled behavior occurs within a non-mockable method which is called by the MUT, and is thus indirectly called by the generated test. Our results demonstrate that automatic mocking is full of caveats and handling all corner cases is an important direction for future work on automated mock generation.}

\begin{mdframed}[style=mpdframe,nobreak=true,frametitle=Answer to RQ3]
\rick succeeds in generating $294$ tests, of which $154$ ($52.4\%$) fully mimic production observations with fully passing oracles. For $19.4\%$ of the test cases, at least one oracle statement fails, showing that the oracles can indeed differentiate between successfully mimicked and incompletely mimicked contexts. 
At runtime, the majority of the tests generated by \rick completely mimic production behavior in the sense that the state asserted by the oracle is equal to the one observed in production.
The cases where the generated tests fail at runtime reveal promising research directions for sophisticated production monitoring tools, such as effective deserialization and efficient resource snapshotting.
\end{mdframed}

\subsection{Results for RQ4 [Effectiveness]}\label{sec:results-rq4}

\begin{figure*}[t!]
    \centering
    \begin{subfigure}[t]{0.33\textwidth}
        \centering
        \includegraphics[height=2.2in]{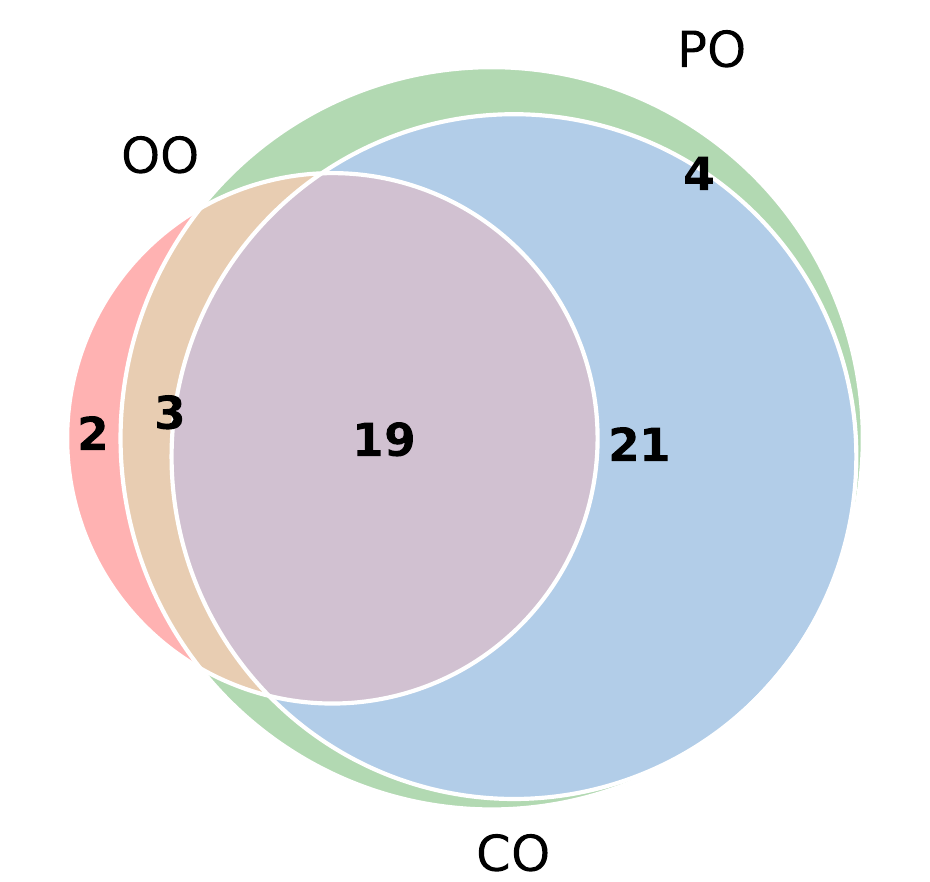}
        \caption{\graphhopper}
        \label{fig:venn-graphhopper}
    \end{subfigure}%
    ~ 
    \begin{subfigure}[t]{0.33\textwidth}
        \centering
        \includegraphics[height=2.2in]{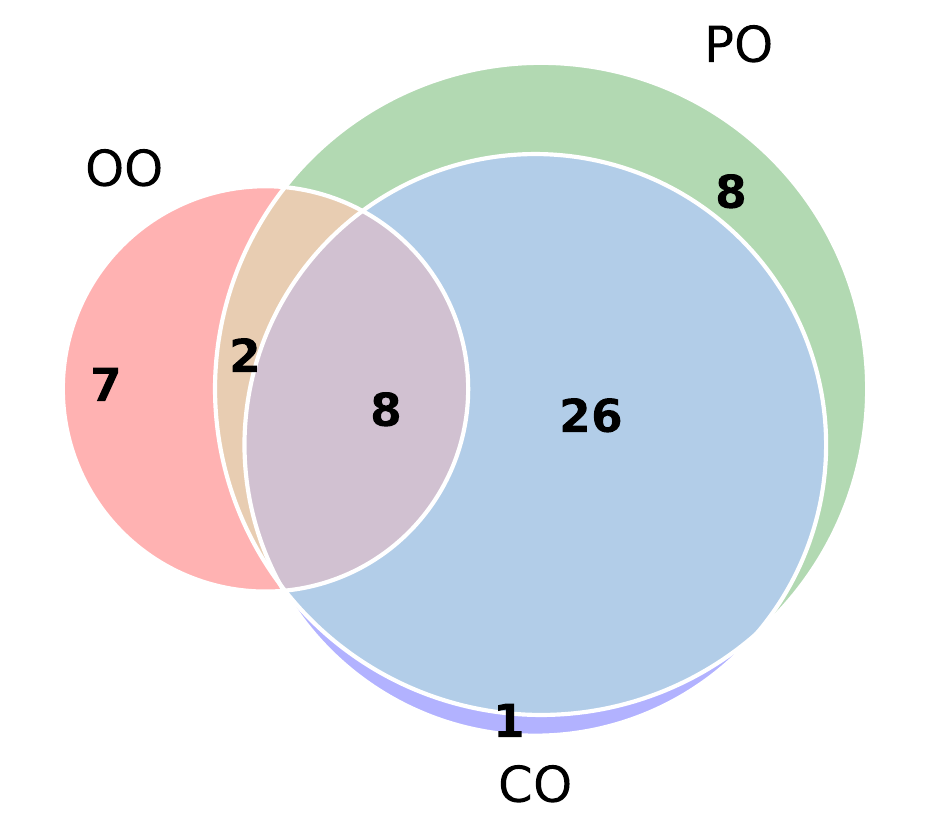}
        \caption{\gephi}
        \label{fig:venn-gephi}
    \end{subfigure}%
    ~ 
    \begin{subfigure}[t]{0.33\textwidth}
        \centering
        \includegraphics[height=2.2in]{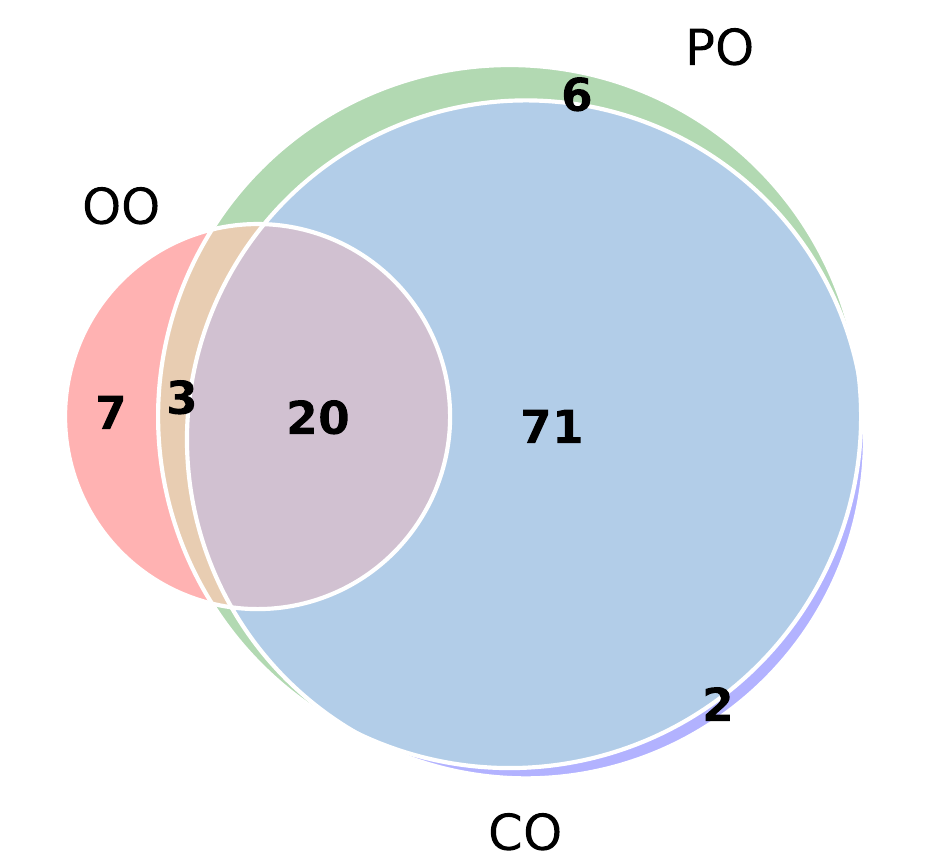}
        \caption{\pdfbox}
        \label{fig:venn-pdfbox}
    \end{subfigure}
    \caption{The \rick tests kill 49 mutants in \graphhopper, 52 mutants in \gephi, and 109 mutants in \pdfbox. The three types of mock-based oracles complement each other to detect regressions in all three projects.}
    \label{fig:mutation-venn}
\end{figure*}

{As described in \autoref{sec:protocol}, we want to determine the effectiveness of \rick tests at determining regressions \cite{just2014mutants}.
We generate a set of first-order mutants for each MUT with LittleDarwin \cite{parsai2017littledarwin}.
We focus the generation of mutants for the $68$ MUTs across the three projects that have at least one passing \rick test.
Furthermore, we consider the mutants that are covered by the test input in the generated tests, i.e., mutants that lie on the path of the MUTs exercised by the tests.
This is because a mutant that lies on an uncovered path will be undetectable by design \cite{petrovic2021practical}.
In total, we consider $449$ mutants: $69$ mutants for the $14$ \graphhopper MUTs that have at least one passing test, $107$ mutants for the $21$ MUTs in \gephi, and $273$ mutants for the $33$ \pdfbox MUTs.
Our replication package\footnote{\url{https://github.com/ASSERT-KTH/rick-experiments}} contains the automated mutation analysis pipeline, as well as the generated mutants and test execution logs.
We also include detailed reports on the set of mutants detected by each mock-based oracle of each MUT.}

{
Our findings from the execution of the generated tests against the mutants are summarized in \autoref{fig:mutation-venn}.
The Venn diagrams represent the distribution of the $210$ mutants killed by \OO, \PO, and \CO for the three case studies.
We note from the Venn diagrams that $19$ mutants in \graphhopper, $8$ mutants in \gephi, and $20$ in \pdfbox are killed by all three mock-based oracles.
Meanwhile, for all three projects, the three mock-based oracles differ in their ability to detect mutants.
For example, per \autoref{fig:venn-graphhopper}, $2$ and $4$ mutants in \graphhopper are detected only by \OO and \PO, respectively.
Likewise, in \autoref{fig:venn-pdfbox}, $2$ mutants in \pdfbox result in extra invocations of a mocked method that are only detected by \CO.
Moreover, in \autoref{fig:venn-gephi}, $2$ mutants in \gephi are killed by \OO and \PO, but not \CO.
Across the three projects, \OO kills $16$ mutants, \PO kills $18$ mutants, and \CO kills $2$ mutants that are undetected by other oracles.
The lower number of mutants killed only by \CO can be attributed to the fact that LittleDarwin does not include a mutation operator that directly removes method calls.
Also, mocked methods may be invoked in the expected order and frequency, but with different, mutated parameters.
This will not be detected by \CO but will be detected by \PO.
The set of mutants killed by \OO is always smaller than for the other oracles.
This is because, as highlighted in \autoref{sec:results-rq1}, we generate \OO tests only for MUTs that return primitive values.
Yet, when present, \OO kills mutants in all three projects, i.e., $24$ mutants in \graphhopper, $17$ in \gephi, and $30$ mutants in \pdfbox.}

{
Overall, all three types of oracles are effective at detecting regressions.
We have also observed that the generated tests kill at least one mutant for each MUT.
These observations are evidence that the \rick tests, with inputs sourced from production, indeed specify the behavior of the MUTs.
Moreover, \OO, \PO, and \CO can detect different bugs.
The \rick tests with mock-based oracles can therefore complement each other, even given the same test input.
This aligns with the findings of Staats \textit{et al.} \cite{StaatsGH12} and Gay \textit{et al.} \cite{GaySWH15a} that multiple oracles specified for a test input may perform differently with respect to their fault-finding ability.
}

\begin{lstlisting}[language=Java, belowskip={-10pt}, label={lst:loadExisting-original}, caption={The original MUT\#8 in \graphhopper, \texttt{loadExisting}}, float]
public class LineIntIndex {
  ...
  public boolean loadExisting() {
    ...
    if (!dataAccess.loadExisting())
      return false;
    ...
    GHUtility.checkDAVersion(..., dataAccess.getHeader(0));
    checksum = dataAccess.getHeader(1 * 4);
    minResolutionInMeter = dataAccess.getHeader(2 * 4);
    ...
    return true;
  }
}

\end{lstlisting}

\begin{lstlisting}[language=diff, belowskip={-10pt}, numbers=none, label={lst:loadExisting-mutants}, caption={Five first-order mutants are introduced in \texttt{loadExisting}. The line numbers correspond to the line numbers in \autoref{lst:loadExisting-original}.}, float]
// Mutant #1: Extreme mutation - Lines 4 to 12
public boolean loadExisting() {
-  if (initialized)
-  ...
-  return true;
+  return true;
}

// Mutant #2: Extreme mutation - Lines 4 to 12
public boolean loadExisting() {
-  if (initialized)
-  ...
-  return true;
+  return false;
}

// Mutant #3: Line 5
- if (!dataAccess.loadExisting())
+ if (dataAccess.loadExisting())

// Mutant #4: Line 9
- checksum = dataAccess.getHeader(1 * 4);
+ checksum = dataAccess.getHeader(1 / 4);

// Mutant #5: Line 10
- minResolutionInMeter = dataAccess.getHeader(2 * 4);
+ minResolutionInMeter = dataAccess.getHeader(2 / 4);

\end{lstlisting}

\begin{lstlisting}[language=diff, belowskip={-10pt}, label={lst:loadExisting-test-template}, caption={The \emph{Arrange} and \emph{Act} phases of the \rick tests for \texttt{loadExisting}}, float]
@Test
@DisplayName("Test for loadExisting, mocking
DataAccess.loadExisting(), DataAccess.getHeader(int)")
public void testLoadExisting() {
  // Arrange
  LineIntIndex receivingObject = deserialize( "receiving.xml");
  DataAccess mockDataAccess = insertMockField_DataAccess_InLineIntIndex( receivingObject);
  when(mockDataAccess.loadExisting()).thenReturn(true);
  when(mockDataAccess.getHeader(0)).thenReturn(5);
  when(mockDataAccess.getHeader(4)).thenReturn(1813699);
  when(mockDataAccess.getHeader(8)).thenReturn(300);
  
  // Act
  boolean actual = receivingObject.loadExisting();

  // Assert
  ...
\end{lstlisting}

{
We illustrate this phenomenon using the example of MUT\#8 in \graphhopper, which is the \texttt{loadExisting} method presented in \autoref{lst:loadExisting-original}.
\autoref{lst:loadExisting-test-template} presents the common \emph{Arrange} and \emph{Act} phases of the three tests generated by \rick for \texttt{loadExisting}.
\autoref{lst:loadExisting-oo}, \autoref{lst:loadExisting-po}, and \autoref{lst:loadExisting-co} contain the \emph{Assert} phase of the \OO, \PO, and \CO test, respectively.
The $5$ mutants produced by LittleDarwin for \texttt{loadExisting}, are shown in \autoref{lst:loadExisting-mutants}.
The generated \OO test detects $2$ of these mutants (\#2 and \#3), as the assertion (\autoref{lst:loadExisting-oo}) fails on an output that differs from the expected \texttt{boolean} value.
The \CO (\autoref{lst:loadExisting-co}) kills $3$ mutants (\#1, \#2, and \#3), as the invocations to the methods \texttt{loadExisting} and \texttt{getHeader} are expected on the mocked \texttt{DataAccess} object, but do not occur due to the mutation.
The verification statements in the \PO (\autoref{lst:loadExisting-po}) kill all $5$ mutants.
This is because the expected mock method invocations within the MUT either do not occur entirely, or occur with unexpected parameters, causing the \PO test to fail.
}

\begin{lstlisting}[language=Java, belowskip={-10pt}, label={lst:loadExisting-oo}, caption={The generated \OO for \texttt{loadExisting}}, firstnumber=17, float]
  assertEquals(true, actual);
}
\end{lstlisting}

\begin{lstlisting}[language=Java, belowskip={-10pt}, label={lst:loadExisting-po}, caption={The generated \PO for \texttt{loadExisting}}, firstnumber=17, float]
  verify(mockDataAccess, atLeastOnce()).loadExisting();
  verify(mockDataAccess, atLeastOnce()).getHeader(0);
  verify(mockDataAccess, atLeastOnce()).getHeader(4);
  verify(mockDataAccess, atLeastOnce()).getHeader(8);
}
\end{lstlisting}

\begin{lstlisting}[language=Java, belowskip={-10pt}, label={lst:loadExisting-co}, caption={The generated \CO for \texttt{loadExisting}}, firstnumber=17, float]
  InOrder orderVerifier = inOrder(mockDataAccess);
  orderVerifier.verify(mockDataAccess, times(1)).loadExisting();
  orderVerifier.verify(mockDataAccess, times(3)) .getHeader(anyInt());
}
\end{lstlisting}

{
We now discuss cases where mutants are not killed by \rick tests.
First, a mutant will be undetected if it results in a behavior that is equivalent to that of the original MUT. This is a well-known limitation of mutation analysis \cite{madeyski2013overcoming}.
\revisedthree{Second, a mutant will be undetected if the production input cannot infect the program's state, or if a mock-based oracle does not capture its side-effects.
The challenges of producing test cases that can effectively infect, propagate and observe the effects of a mutant are well known in the literature  \cite{du2024ripples,veraperez2019}.
An interesting prospect for future work is to trigger the detection of alive covered mutants through stubbing.
}
For example, we observe this for mutant \#1 in \texttt{loadExisting} (\autoref{lst:loadExisting-mutants}), which is undetected by the \OO in \autoref{lst:loadExisting-oo}.
The mutation causes the MUT to directly return \texttt{true}, which is indeed the expected value specified by the \OO.
However, this mutant is killed by \PO and \CO, as the mock method calls they specify no longer occur within the MUT.
Similarly, the \CO in \autoref{lst:loadExisting-co} does not kill mutants \#4 and \#5 in \autoref{lst:loadExisting-mutants}.
The method calls specified by the \CO still occur with the same frequency and in the expected order, but with different parameters.
Parameter verification is not the focus of the \CO, but the \PO in \autoref{lst:loadExisting-po} detects the mutants and fails.
}

\begin{mdframed}[style=mpdframe,nobreak=true,frametitle=Answer to RQ4]
{
All three types of oracles are effective at detecting regressions introduced within MUTs. 
Moreover, $16$, $18$, and $2$ mutants are only killed by \OO, \PO, and \CO, respectively.
The \rick tests with mock-based oracles  complement each other for regression testing, making them a useful addition to the test suite. 
}
\end{mdframed}

\subsection{Results for RQ5 [Quality]}\label{sec:results-rq5}
\begin{table*}
\renewcommand*{\arraystretch}{2}
\centering
\caption{Profiles of the developers who participated in the survey to assess the quality of tests generated by \rick}\label{tab:participant-profiles}
\begin{tabular}{@{}|l|r|r|r|r|r|r@{}}
\hline
\textbf{\textsc{Participant}} & 
{\begin{tabular}[c]{@{}c@{}}\textbf{\textsc{Experience}}\\ \textbf{\textsc{(Years)}} \end{tabular}} & \textbf{\textsc{Sector}} & \textbf{\textsc{Writes Tests}} & \textbf{\textsc{Uses Mocks}} &
{\begin{tabular}[c]{@{}c@{}}\textbf{\textsc{Programming}}\\ \textbf{\textsc{Environment}} \end{tabular}} \\
\hline
\textsc{P1} & 6 & Consultancy & \emph{``everyday, testing is [my] life"} & sometimes & \texttt{JUnit + Mockito} \\ \hline
\textsc{P2} & 30 & Consultancy  & often & sometimes & \texttt{JUnit + Mockito} \\ \hline
\textsc{P3} & 7 & Telecom & TDD practitioner & sometimes & \texttt{Python (pytest)} \\ \hline
\textsc{P4} & 10 & Product (\graphhopper) & \emph{``all the time"} & sometimes &  \texttt{JUnit + Mockito} \\ \hline
\textsc{P5} & 14 & Game development & sometimes & rarely, \emph{``want to mock more``} & \texttt{C, C\#, .NET} \\ \hline
\end{tabular}
\end{table*}

Per the protocol described in \autoref{sec:protocol}, we have interviewed $5$ developers between June and July, 2022, with the goal of assessing their opinion on the tests generated by \rick.
\autoref{tab:participant-profiles} presents the details of the participants of the survey. The five developers work in different sectors of the IT industry, and have between $6$ and $30$ years of experience with software development. Notably, participant P4 is a core contributor to \graphhopper, with the highest number of commits to its GitHub repository in the last 5 years. From \autoref{tab:participant-profiles}, we see that all participants write tests sometimes or everyday, and most of them also define and use mocks. The developers also work with diverse programming environments. P1, P2, and P4 work with Java testing and mocking frameworks, specifically JUnit and Mockito. P3 is a Python developer, while P5, who works in a game development company, works mostly with C, C\#, and .NET framework.

We begin each meeting by introducing the concepts and terminology of mock-based testing, the \rick pipeline, and our experiments with \graphhopper.
\revisedfour{We demonstrate the features of the tests generated by \rick by selecting a total of $6$ generated tests, one for each of the three mock-based oracles for $2$ MUTs defined in \graphhopper. 
We select the first two MUTs from \autoref{tab:graphhopper-results} that meet the selection criteria mentioned in \autoref{sec:protocol}.
The two selected MUTs, MUT\#16 and MUT\#6, have $58$ and $13$ LOC, respectively.
The tests generated for MUT\#16 have two stubs, and a method call on an external parameter mock object.
The tests for MUT\#6 have one stub and three method calls on a mocked field.
Excluding comments, the number of lines of code across the six generated tests is 39 (median 6 lines of code for each test).
}
We introduce the two MUTs, MUT\#6 and MUT\#16, to the participant, also presenting their source code. We invite them to clone a fork of \graphhopper which includes the generated tests. During the meeting, we browse through the generated tests with them via screen sharing.
Finally, we ask the participant three sets of questions about the generated tests while documenting their responses. These questions relate to \emph{mocking effectiveness}, i.e., how mocks are used within the tests, as well as the \emph{structure} and \emph{understandability} of the generated tests.

\emph{Mocking effectiveness}: The first set of questions relates to how the mocks are used in the generated tests. Per their answers, all five participants agree that the generated tests for the 2 MUTs represent realistic behavior of \graphhopper in production, which would be useful for developers. P2 observed that this can \emph{``save the time spent on deciding the combination of inputs and finding corner cases, especially for methods with branches."} P5 added that, according to them, collecting data from production is \emph{``where \rick shines most, since it abstracts away [for developers] the tricky exercise of deciding test inputs and internal states."} Furthermore, we had detailed discussions with the participants about the verification statements in \textbf{PO} and \textbf{CO} tests. P1, P3, and P5 noted that they contribute differently to the verification of the behavior of the MUT. P5 remarked that while some verification statements may be redundant, they \emph{``can be manually customized by developers"} when generated tests are presented to them. However, P3, who works primarily with Python, a dynamically-typed language, commented that for them \emph{``the verification of the frequency of the mock method calls in the CO tests is useful, but not the \texttt{anyInt()} or \texttt{anyString()} wildcards to match argument types."} Additionally, P3 and P4 also discussed the stability of these tests with respect to code refactoring, with P4 saying that \emph{``the tests might break in case a developer refactors legacy code. But because they are so detailed, a regression can also be figured out at a very low level."} P5 highlighted an interesting aspect about the human element in software development by sharing that \emph{``while \rick fits exactly an actual problem in the industry, a potential disadvantage is that it can spoil developers who may become incentivized to design without testability in mind. The tests can be automatically produced later when the application is production-ready."}

\emph{Structure}: Next, we assess the opinion of developers about the structure of the generated tests. All $5$ developers appreciated the ``Arrange-Act-Assert" pattern \cite{wei2022automatically} that is systematically followed in the generated tests. They note that this pattern makes the structure of the tests clear. P2, who has experimented with other test generation tools, noted that clear structure and intention is important to improve the adoption of automated tools. All the developers mentioned that they typically use the pattern when writing tests, including P5 who added that the structure was \emph{``spot on."}
Moreover, P1, P2, and P4 were appreciative of the description generated by \rick for each test, using the \texttt{@DisplayName} JUnit annotation, with P2 noting that \emph{``it is rare and useful."}

\emph{Understandability}: Finally, we question each participant about the understandability of the generated tests. P1 and P2 mentioned that the comments demarcating each phase in the generated tests contribute to their intuitiveness and make their intention clearer, with P5 exclaiming that they make the tests \emph{``super easy to visualize."} Additionally, P2, P4, and P5 noted that the comments are useful, especially since they help understand what is happening within tests generated automatically. However, P4 observed that while the comments \emph{``do not hurt"}, they would not add them while writing the test manually. P3 was also of the opinion that the comments may be removed without impacting the  understandability of the tests.

The perception of developers of automatically generated tests and mocks, using data collected from production, is valuable qualitative feedback about the relevance of \rick. 
The $5$ experienced developers and testers confirm that the data collected in production is realistic and useful to generate tests and mocks. They also appreciate the systematic structure of the test cases, as well as the explicit intention documented in comments.

This qualitative study also suggests the need for further work in the area of automated mock generation.
For example, P3 expressed interest in analyzing the influence of the architecture of the system under test on the stability of the mocks. P4 observed that it would be useful to evaluate the overall testability of an application in terms of how many MUTs have mockable method calls. We also discussed with P5 about adding more context and business-awareness to test names and comments to further help developers troubleshoot test executions. We identify these as excellent directions for further work, with much potential for impact on the industry.

{
Furthermore, from our discussions with the developers, we find that they all agree that mocking is advantageous.
However, developers often rely on metrics such as coverage as a proxy for the strength of their test suite. This leads to a methodological mismatch where mocks are desirable, yet do not contribute directly to the strength of the tests, i.e., do not have an impact on test coverage. 
We note that an implicit prerequisite for developers to be more open to the benefits of using mocks is to consider test quality beyond coverage, to embrace the value of mocking, as well as the effort required to include mocks in their pipeline.
Mocking is not trivial and comes with the challenges highlighted in \autoref{sec:mocking-challenges}.
\rick can help with realistic mocks, directly available in readable tests, that can capture regressions.}

\begin{mdframed}[style=mpdframe,nobreak=true,frametitle=Answer to RQ5]
Five experienced developers confirm that the concept of data collection in production is relevant for the generation of tests with mocks. They all appreciate the systematic ``Arrange-Act-Assert" template for the test which contributes to the overall good understandability of the  tests generated by \rick. 
\end{mdframed}

\section{Discussion}\label{sec:discussion}
\revisedthree{We now discuss the performance of \rick, the limitations of our approach for mock generation, and the threats to the validity of our findings.}

\subsection{Performance}\label{sec:performance}
\revisedthree{
Runtime data capturing is a key process within \rick, which requires some additional computation and memory resources.
We adapt the methodology used in previous work \cite{9526340} to measure the performance implications of \rick during the execution of our three case studies.
We exercise each application with the workload described in \autoref{sec:production-usage} in three distinct ways.
First, we determine the baseline performance of the application by running it without any agent attached.
Next, we run it with the default monitoring agent, i.e., Glowroot, attached (recall that Glowroot is standard monitoring technology used in the industry).
Finally, we attach \rick to the application as a Glowroot plugin, which means the complete monitoring plus data collection machinery for the MUT and mockable method invocations.
The experiments are performed on a machine running Ubuntu 22.04, with an 8-core Intel i5 processor and 16GB memory.
We use the Linux \texttt{top} command, filtered on the application name, to obtain its CPU and memory usage.}

\revisedthree{
For \graphhopper, the average CPU consumption for the baseline execution is 35.4\% while the memory usage is 824.7 MB.
Attaching Glowroot as a monitoring agent to \graphhopper increases the CPU and memory consumption to 66\% and 983.4 MB.
Attaching the complete monitoring and data capturing abilities of \rick results in the CPU and memory usage of 109.2\% and 1570.2 MB.
This means that the execution with \rick and all MUTs and mockable methods monitored, consumes thrice the CPU compared to baseline, and twice the memory.
Next, normal execution of \gephi consumes 117.6\% CPU and 856.5 MB memory on average.
Monitoring with Glowroot increases these usages to 142.2\% and 1157.8 MB, respectively.
Attaching \rick with \gephi results in 249.1\% CPU and 1601.9 MB memory usage.
The resource consumption during execution of \gephi with \rick is about twice the baseline amount.
Finally, averaging across 10 executions of \pdfbox, we find that its CPU consumption is 92\%, while its memory usage is 63.4 MB.
Attaching Glowroot increases CPU and memory consumption to 335.4\% and 190.3 MB, respectively.
Attaching \rick does not contribute to additional CPU consumption, and leads to an increased memory usage at 428.2 MB (about 6.7 times the baseline).
Overall, we note that monitoring contributes to additional resource consumption for all cases, with respect to the baseline.
This increases more with \rick as a consequence of dynamic instrumentation and serialization.
}

\revisedthree{
Monitoring is an essential component of modern observability, for ensuring smooth operation and diagnosis \cite{maguire2024automation}.
\rick leverages monitoring to generate unit tests that reflect production behaviors and detect regressions.
However, monitoring comes at a cost, impacting the scalability of \rick, as well as similar approaches for observation-based test generation \cite{9526340, alshahwan2024observation}.
We have taken measures to mitigate this issue, designing \rick to be configurable.
First, developers configure the target methods to monitor and capture data for, and only those are instrumented.
Additionally, they also specify the number of invocations of these methods that must be monitored in production.
Being a research prototype, it is clear that the monitoring agent has ample room for performance optimization when productized.
Furthermore, as we highlight in \autoref{sec:rick-sdlc}, \rick is only meant to be periodically  employed, for every testing campaign when the development team focuses on improving unit tests with production data.}

\revisedfour{\subsection{Subsequent Test Failures}
As we demonstrate with RQ4, the potential future failures of a test generated by \rick are meant to be indicative of regressions in the method under test, compared to the current behavior.
However, there are two cases when a \rick test would fail in the absence of a regression.}

\revisedfour{First, for mocked methods that perform non-deterministic actions, such as network calls, the stubs and mock-based oracles generated by \rick reflect the responses observed in the field, by design.
For instance, calls that result in valid responses (2XX) or 4XX errors, such as a 404 for a non-existent resource, will lead to the generation of useful tests with \rick.
However, \rick will generate a test that may overfit a situation with a 5XX error.
Such a response might not be reproduced within the generated test, causing it to fail.}

\begin{lstlisting}[belowskip={-10pt}, label={lst:pre-refactoring}, caption={\rick generates tests with the three mock-based oracles for MUT\#27 of \pdfbox, \texttt{getWidthFromFont}.
Refactoring \texttt{getWidthFromFont} can impact the validity of these oracles.}, float]
public float getWidthFromFont(int code) {
  int gid = codeToGID(code);
  float width = ttf.getAdvanceWidth(gid);
  float unitsPerEM = ttf.getUnitsPerEm();
  if (unitsPerEM != 1000) {
    width *= 1000f / unitsPerEM;
  }
  return width;
}%@\par\noindent\dotfill@)
@Test
public void testGetWidthFromFont_OO() {
  ...
  // Assert
  assertEquals(750.0, actual, 0.0);
}%@\par\noindent\dotfill@)
@Test
public void testGetWidthFromFont_PO() {
  ...
  // Assert
  verify(mockTTF, atLeastOnce()).getAdvanceWidth(0);
  verify(mockTTF, atLeastOnce()).getUnitsPerEm();
}%@\par\noindent\dotfill@)
@Test
public void testGetWidthFromFont_CO() {
  ...
  // Assert
  InOrder ordVerifier = inOrder(mockTTF);
  ordVerifier.verify(mockTTF, times(1)).getAdvanceWidth(anyInt());
  ordVerifier.verify(mockTTF, times(1)).getUnitsPerEm();
}
\end{lstlisting}
\begin{lstlisting}[belowskip={-10pt}, label={lst:post-refactoring}, caption={MUT\#27 of \pdfbox, \texttt{getWidthFromFont}, after refactoring.
Relative to \autoref{lst:pre-refactoring}, its mockable method calls \texttt{getAdvanceWidth} and \texttt{getUnitsPerEm} are reordered, and the former is renamed to \texttt{calculateAdvanceWidth}.}, float]
// Refactored getWidthFromFont
public float getWidthFromFont(int code) {
  int gid = codeToGID(code);
  float unitsPerEM = ttf.getUnitsPerEm(); // reordered call
  float width = ttf.calculateAdvanceWidth(gid); // renamed method
  if (unitsPerEM != 1000) {
    width *= 1000f / unitsPerEM;
  }
  return width;
}
\end{lstlisting}

\revisedfour{Second, the failure of a \rick-generated test may be the result of a refactoring \cite{kim2014empirical}, and not a regression.
For example, for a method under test \emph{m}, the call oracle will fail if the order of the mock method calls within \emph{m} is changed, while the semantics of \emph{m} is preserved.
Also, the parameter oracle may not hold if the arguments received by at least one mock method call within \emph{m} deviate from their expected values.
This can indicate a behavioral change within \emph{m}, but also a refactoring of the stubbed method call.
The parameter and call oracles are sensitive to refactoring changes.
However, a failing output oracle implies that the output from \emph{m} is unequal to the expected output, signifying a behavioral change in the public API.}

\revisedfive{Consider lines 1 to 9 of \autoref{lst:pre-refactoring}, which present the method \texttt{getWidthFromFont} of \pdfbox (MUT\#27 in \autoref{tab:pdfbox-results}).
This MUT calls two mock methods on the field \texttt{ttf} of type \texttt{TrueTypeFont}, \texttt{getAdvanceWidth} (line 4) and \texttt{getUnitsPerEm} (line 5).
The mock-based oracles generated by \rick for \texttt{getWidthFromFont} are presented on line 15 (\textbf{OO}), lines 22-23 (\textbf{PO}), and lines 30-32 (\textbf{CO}). 
\autoref{lst:post-refactoring} shows a refactored version of \texttt{getWidthFromFont}, with two semantically-preserving changes: 1) the mock method calls are reordered, i.e., \texttt{getUnitsPerEm} occurs first (line 4), and 2) \texttt{getAdvanceWidth} is renamed to \texttt{calculateAdvanceWidth} (line 5).
These changes cause a failure of the \textbf{CO} and \textbf{PO}, but the \textbf{OO} still holds.}

\revisedfive{Despite their varying degree of sensitivity, all three mock-based oracles alert the developer of behavioral changes that are introduced in their code, which is also the goal of testing. 
The impact of code refactoring on mock-based oracles is an important direction for future work.
}

\subsection{Threats \& Limitations}

\revisedthree{\textbf{Serialization}
The internal validity of \rick is impacted by technical limitations.
For example, instrumentation of some methods may fail \cite{wachter2024serializing}.
One of the biggest technical challenge is the serialization and deserialization of large and complex objects.
This may result in incomplete or unfaithful program states within the generated tests, which do not reflect the ones observed in production.
}

\noindent\revisedthree{\textbf{Application Vs Library} 
A source of threat to the external validity of our findings arises from the software projects we consider for the evaluation of \rick.
We make sure that the three projects are 1) complex, and 2) from diverse domains.
Indeed, we consider a library, a desktop application, and a backend application.
We do not guarantee that our findings hold for applications and libraries in other languages such as TypeScript, or nim.
Still, we believe that \rick would not let them down.}

\noindent\revisedthree{\textbf{Program Evolution} 
Tests and programs co-evolve within software projects \cite{shimmi2022leveraging,marsavina2014studying,le2021untangling}. 
The tests generated by \rick are no different, they may be impacted by changes in the application code.
A refactoring in the source code, such as a modification in the order of method invocations, or a change in the parameters of an existing method, may require changes in the generated tests.
However, if these modifications are not semantically relevant, this is tedious, low value work for developers.
This limitation is shared by all regression test generation techniques.
In this case, developers can always regenerate new tests when significant changes are made to the methods under tests.
}

\section{Related Work}\label{sec:related-work}

This section presents the literature on mock objects, as well as their automated generation. We also discuss studies about the use of information collected from production for the generation of tests.

\subsection{Studies on Mocking}
Since mocks were first proposed \cite{mackinnon2000endo}, they have been widely studied \cite{thomas2002mock, freeman2004mock}. 
Their use has been analyzed for major platforms, such as Java \cite{spadini2019mock}, Python \cite{7927976}, C \cite{Mudduluru534312}, Scala \cite{laufer2019tests}, Android \cite{fazzini2020framework, 9794020}, PHP, and JavaScript \cite{de2023mock}.
These studies highlight the prevalence and practices of defining and using mocks.
Some empirical studies analyze more specific aspects about the usage of mocks, such as their definition through developer-written mock classes \cite{9240675}, or mocking frameworks \cite{6958396}, or their use in the replacement of calls to the file system \cite{marri2009empirical}.
\revisedthree{Xiao \textit{et al.} \cite{xiao4100265empirical} find that mocking is practiced in 66\% of the 264 projects of the Apache Software Foundation.}
Spadini \textit{et al.} discuss the criteria developers consider when deciding what to mock \cite{spadini2017mock}, as well as how these mocks evolve \cite{spadini2019mock}.
MockSniffer by Zhu \textit{et al.} \cite{10.1145/3324884.3416539} uses machine learning models to recommend mocks.
Mockingbird by Lockwood \textit{et al.} \cite{lockwood2019mockingbird} uses mocks to isolate the code under dynamic analysis from its dependencies.
The use of mocks for Test Driven Development \cite{kim2006mock}, modeling \cite{stoel2021modeling}, and as an educational tool for object-oriented design \cite{nandigam2009interface, nandigam2010using} has also been investigated.
These works have been inspirational for us in many respects.
Moreover, \rick is designed to generate tests that use Mockito, which is the most popular mocking framework \cite{6958396, de2023mock, xiao4100265empirical}, and is itself a subject of study \cite{gay2016challenges, turner2016multi, wang2019attention, kim2021studying}.
However, none of these related works touch upon mocking in the context of production monitoring, specifically generating mocks and mock-based oracles.
These are the two key contributions of our work.

\subsection{Mock Generation}
Several studies propose approaches to automatically generate mocks, albeit not from production executions.
For example, search-based test generation can be extended to include mock objects \cite{ArcuriFJ17}, to mock calls to the file system \cite{arcuri2014automated} and the network \cite{arcuri2015generating}. Symbolic execution may also be used to generate mocks \cite{tillmann2006mock, alshahwan_et_al:DagSemProc.10111.3, islam2010dsc+}, to mock the file system \cite{5070712}, or a database for use in tests \cite{taneja2010moda}.
Honfi and Micskei \cite{honfi2020automated} generate mocks to replace external interactions with calls to a parameterized sandbox. This sandbox receives inputs from the white-box test generator, Pex \cite{tillmann2008pex}. Moles by Halleux and Tillmann \cite{halleux2010moles} also works with Pex to isolate the unit under test from interactions with dependencies by delegating them to alternative implementations.
Salva and Blot \cite{salva2020using} propose a model-based approach for mock generation. Bhagya \textit{et al.} \cite{bhagya2019generating} use machine learning models to mock HTTP services using network traffic data.
GenUTest \cite{pasternak2009genutest} generates JUnit tests and mock aspects by capturing the interactions that occur between objects during the execution of medium-sized Java programs.
StubCoder \cite{zhu2023stubcoder} by Zhu \textit{et al.} uses an evolutionary algorithm to generate new stubs and repair incorrect stubs within existing JUnit tests.
\revisedfour{On the other hand, ARUS \cite{li2024automatically} utilizes information from the execution of the test suite to detect and remove unnecessary stubbing.}
\revisedthree{
Abdi and Demeyer \cite{abdi2022test} leverage mocking within their proposed test transplantation technique that ports client tests into library test suites.
ARTISAN by Gambi \textit{et al.} \cite{gambi2023action} instruments end-to-end GUI tests of Android applications, in order to carve unit tests that mock classes of the Android framework.
}

To the best of our knowledge, \rick is the only tool that generates mock-based oracles to verify the behavior of the system under test, per the the production executions, with real usages and real data.

The definition and behavior of mocks can also be extracted from other artifacts. For instance, the design contract of the type being mocked can be used to define the behavior of a mock \cite{galler2010automatically}. Samimi \textit{et al.} \cite{samimi2013declarative} propose declarative mocking, an approach that uses constraint solving with executable specifications of the mock method calls. 
Solms and Marshall \cite{solms2016contract} extract the behavior of mock objects from interfaces that specify their contract.
Wang \textit{et al.} \cite{wang2021automatic} propose an approach to refactor out developer-written subclasses that are used for the purpose of mocking, and replace them with Mockito mocks.
Mocks have also been generated in the context of cloud computing \cite{zhang2011environmental}, such as for the emulation of infrastructure by MockFog \cite{9411706}. Jacinto \textit{et al.} \cite{jacinto2020test} propose a mock-testing mode for drag-and-drop application development platforms.
Contrary to these approaches, \rick monitors applications in production in order to generate mocks. Consequently, the generated tests reflect the behavior of an application with respect to actual user interactions.

The executions of system tests in the existing test suite can also be leveraged to generate mocks. This approach has been used by Saff \textit{et al.} \cite{saff2005automatic} through system test executions, and Fazzini \textit{et al.} \cite{fazzini2020framework} to generate mocks for mobile applications. 
Bragg \textit{et al.} \cite{bragg2021program} use the test suite of Sketch programs to generate mocks in order to modularize program synthesis.
{However, system tests are artifacts written by developers, and can therefore suffer from biases that developers have about how the system should behave. In contrast, production executions, where \rick sources its test inputs, are free from these assumptions, and reflect how the system actually behaves under real workloads.}

\subsection{Capture and Replay}
Many studies propose techniques to capture a sequence of events that occur within an executing system, with the goal of replaying it \cite{joshi2007scarpe}.
The premise of these techniques is to replicate the state of the system as it was at a certain point in time.
Capture and replay has been successfully applied for the reproduction of crashes \cite{roehm2013monitoring} and failures \cite{bell2013chronicler} that occur in the field.
The captured sequence of events leading up to a crash or failure allows for more efficient debugging when replayed offline by developers \cite{jin2012bugredux, burg2013interactive}, as well as the evaluation of candidate patches for bugs \cite{saieva2022update}.
Capture and replay can also be used to exercise the same sequence of interactions with an application GUI as was done by end-users \cite{steven2000jrapture}.
This can be used to analyze the performance of interactive applications \cite{adamoli2011automated}.
All these existing techniques do not generate an explicit oracle.
They instead rely on an implicit oracle, such as the reproduction of a failure. 
\revisedthree{Saff \textit{et al.} \cite{saff2005automatic} carve focused unit tests from system tests.
Their technique cannot be applied to production environments without major challenges.
In particular, a key challenge that we address with \rick is the serialization of  production objects with reasonable overhead.
This challenge is also noted by Meta, who propose TestGen for generating observation-based tests for Instagram \cite{alshahwan2024observation}.
This aspect, together with our specific oracles, are fundamentally novel compared to the technique of Saff \textit{et al.} \cite{saff2005automatic}.
In addition, their evaluation considers one program, while our evaluation considers three real-world programs exercised with representative field workloads.
}

{\rick is fundamentally different from capture and replay techniques since it generates full-fledged test cases, which include an explicit oracle in an assertion.
This essential difference allows us to assess the effectiveness of the generated tests with mutation analysis, which none of the capture and replay techniques do.}

\subsection{Production-based Oracles}
Monitoring an executing application with the goal of generating tests is an effective means of bridging the gap between the developers' understanding of their system, and how it is actually exercised by users \cite{wang2017behavioral}. To this end, several studies propose tools that capture runtime information. 
Thummalapenta \textit{et al.} \cite{thummalapenta2010dygen} use execution traces for the generation of parameterized unit tests. Wang and Orso \cite{9240614} capture the sequence of method executions in the field, and apply symbolic execution to generate tests for untested behavior. Jaygarl \textit{et al.} \cite{jaygarl2010ocat} capture objects from program executions, which can then be used as inputs by other tools for the generation of method sequences. Tiwari \textit{et al.} \cite{9526340} generate tests for inadequately tested methods using production object states.
\revisedthree{\textsc{ProDJ} by Wachter \textit{et al.} \cite{wachter2024serializing} focus on the readability of the unit tests generated from production data, incorporating the objects captured at runtime as plain Java code.}
Incoming production requests have also been utilized to produce tests for databases \cite{yan2018snowtrail} and web applications \cite{zetterlund2022harvesting}.
\rick leverages this methodology with the novel and specific goal of generating tests with mock-based oracles that verify the interactions between a method and objects of external types, as they occur in production.

\section{Conclusion}\label{sec:conclusion}
In this paper, we present \rick, a novel approach for generating tests with mocks, using data captured from the observation of applications executing in production.
\rick instruments a set of methods under test, monitors their invocations in production, and captures data about the methods and the mockable method calls. Finally, \rick generates tests using the captured data.
The mock-based oracles within the generated tests verify distinct aspects of the interactions of the method under test with the external object, such as the output of the method (\OO), the parameters with which invocations are made on the external object (\PO), and the sequence of these invocations (\CO).
Our evaluation with three open-source applications demonstrates that \rick never gives up: 
It monitors and transforms observed production behavior into concrete tests (RQ1).
The data collected from production is expressed within these generated tests as complex receiving objects and parameters for the methods, as well stubs and mock-based oracles (RQ2).
When executed, $52.4\%$ of the generated tests successfully mimic the observed production behavior. This means that they recreate the execution context for the method under test, the stubbed behavior is appropriate, and the oracle verifies that the method under test behaves the same way as it did in production (RQ3).
{The three mock-based oracles can detect regressions within the methods under test, and \OO, \PO, and \CO can complement each other in finding bugs (RQ4).}
Furthermore, our qualitative survey with professional software developers reveals that the data and oracle extracted from production by \rick are relevant, and that the systematic structure of \rick tests is understandable (RQ5).  

Overall, we are the first to demonstrate the feasibility of creating tests with mocks directly from production, in other terms to capture production behavior in isolated tests.
Since the generated tests reflect the actual behavior of an application in terms of concrete inputs and oracles, they are valuable for developers to augment manually crafted inputs with ones that are relevant in production. 

Our findings open up several opportunities for more research. 
It would be useful to handle more kinds of interactions of a method under test with its environment, such as all method calls made on an external object within the method under test, in order to achieve further isolation within the generated tests.
\revisedthree{Future work should also consider different choices of mockable method calls, to support mocking types within dependencies.} 
\revisedfive{Additionally, the impact of code refactoring on automatically generated mock-based oracles warrants a detailed analysis.}

\section*{Acknowledgements}
\noindent This work has been partially supported by the Wallenberg Autonomous Systems and Software Program (WASP) funded by the Knut and Alice Wallenberg Foundation, as well as by the Chains project funded by the Swedish Foundation for Strategic Research (SSF).

\balance
\bibliographystyle{IEEEtran}
\bibliography{main}

\end{document}